\documentclass[11pt,a4paper]{article}

\pdfoutput=1 

\usepackage{jinstpub} 
\usepackage{bm}
\usepackage[usenames,dvipsnames]{xcolor}
\usepackage{comment}
\usepackage{siunitx}
\usepackage{lineno}
\definecolor{forestgreen}{rgb}{0.13, 0.55, 0.13}

\title{Design and construction of Xenoscope -- a full-scale vertical demonstrator for the DARWIN observatory}

\author{L.~Baudis,}
\author[*]{Y.~Biondi,}
\author{M.~Galloway,}
\author[*]{F.~Girard,\note[*]{Corresponding author.}}
\author{A.~Manfredini,}
\author{N.~McFadden,}
\author{R.~Peres,}
\author{P.~Sanchez-Lucas,}
\author[*]{and K.~Thieme}

\affiliation{Department of Physics, University of Zurich, Winterthurerstrasse 190, 8057 Zurich, Switzerland }

\emailAdd{yanina.biondi@physik.uzh.ch, frederic.girard@physik.uzh.ch, kevin.thieme@physik.uzh.ch}
\doi{https://doi.org/10.1088/1748-0221/16/08/P08052}

\abstract{
The DARWIN observatory is a proposed next-generation experiment to search for particle dark matter and other rare interactions. It will operate a \SI{50}{t} liquid xenon detector, with \SI{40}{t} in the time projection chamber (TPC). To inform the final detector design and technical choices, a series of technological questions must first be addressed. Here we describe a full-scale demonstrator in the vertical dimension, Xenoscope, with the main goal of achieving electron drift over a \SI{2.6}{m} distance, which is the scale of the DARWIN TPC. We have designed and constructed the facility infrastructure, including the cryostat, cryogenic and purification systems, the xenon storage and recuperation system, as well as the slow control system. We have also designed a xenon purity monitor and the TPC, with the fabrication of the former nearly complete. In a first commissioning run of the facility without an inner detector, we demonstrated the nominal operational reach of Xenoscope and benchmarked the components of the cryogenic and slow control systems, demonstrating reliable and continuous operation of all subsystems over 40 days. The infrastructure is thus ready for the integration of the purity monitor, followed by the TPC. Further applications of the facility include R\&D on the high voltage feedthrough for DARWIN, measurements of electron cloud diffusion, as well as measurements of optical properties of liquid xenon. In the future, Xenoscope will be available as a test platform for the DARWIN collaboration to characterise new detector technologies. 

}

\keywords{Detector design and construction technologies and materials, Gas systems and purification, Noble liquid detectors (scintillation, ionization, double-phase), Time projection chambers}

\arxivnumber{2105.13829} 

\begin{document}

\maketitle
\flushbottom

\section{Introduction}
\label{sec:intro}

DARWIN is a proposed next-generation experiment in low-energy astroparticle physics~\cite{Baudis:2012bc}. It will operate a \SI{40}{tonne} (\SI{50}{tonne} total) liquid xenon (LXe) time projection chamber (TPC) with the main goal of probing the experimentally accessible parameter space for weakly interacting massive particles (WIMPs) as dark matter candidates~\cite{Aalbers:2016jon,Schumann:2015cpa}. With its large target mass and expected ultra-low background level, additional physics goals will be pursued, e.g., the search for the neutrinoless double beta decay of $^{136}$Xe~\cite{Baudis:2013qla,Agostini:2020adk}, the real-time observation of solar $pp$-neutrinos with high statistics~\cite{Baudis:2013qla,Aalbers:2020gsn}, the detection of coherent neutrino-nucleus interactions with solar $^{8}$B neutrinos and from supernovae~\cite{Lang:2016zhv,Raj:2019sci}, as well as searches for solar axions, and for axion-like particles and dark photons as dark matter candidates.

The baseline design of the DARWIN experiment is described in~\cite{Aalbers:2016jon}. The inner detector is a cylindrical, two-phase (liquid/gas) xenon TPC, with both diameter and height of \SI{2.6}{m}, designed for light and charge readout. The TPC is placed in a low-background, double-walled cryostat surrounded  by a neutron veto and a water Cherenkov detector to shield from environmental radioactivity as well as from cosmic muons and their secondary particles. The primary shielding from cosmic rays is achieved by locating the detector deep underground. 

Interactions in the LXe target produce both scintillation and ionisation. A first, prompt light signal ($S1$) is followed by a delayed, proportional scintillation signal ($S2$) from electrons transported upwards by a drift field of around~\SI{200}{V/cm} and extracted into the gas phase by a stronger field. Both light signals, emitted in the vacuum ultra-violet (VUV) range, are detected at the top and bottom of the TPC by photosensor arrays to reconstruct the time, energy and position of an interaction. The time difference between the $S1$ and $S2$ signals yields the depth of an interaction, while the $(x,y)$-coordinates are obtained from the light pattern in the top photosensor array.

The realisation of this large detector requires the demonstration of a series of technologies at the \SI{2.6}{m} drift scale. We therefore designed and built a full-scale demonstrator in the vertical dimension, which we call Xenoscope. The main goal of Xenoscope is to demonstrate electron drift for the first time in a LXe TPC over a \SI{2.6}{m} distance. To this end, two essential aspects must be addressed: the purity of the LXe and the application of high voltage (HV) to obtain an adequate and homogeneous electric drift field.

The purity of the target is quantified by the number of electronegative impurities dissolved in the LXe, which can absorb electrons on their way to the gas phase. For the successful operation of the TPC, the xenon must be continuously recirculated through a purification system. The system must ensure a high purification efficiency coupled to a fast recirculation speed. The strength and homogeneity of the drift field must be optimised for a maximum number of electrons reaching the gas phase, as electrons drift faster when subjected to a stronger electric field. The effect of these two factors is quantified in terms of the electron lifetime $\tau_\mathrm{e}$~\cite{Aprile:2009dv,Chepel:2012sj}, which is the mean time for the charge to drop by a factor of~$e$.
It has been shown that using a liquid recirculation system, which allows for high recirculation speeds ($\sim\SI{1500}{slpm}$), an electron lifetime of \SI{7}{ms} is within reach for systems of several tonnes of LXe~\cite{Macolino:2021}. In Xenoscope, we intend to achieve a uniform drift field of $\sim\SI{200}{V/cm}$, which translates in a drift velocity of $\sim\SI{1.5}{mm/\micro s}$~\cite{Baudis:2017xov}. Therefore, electrons drifting the full \SI{2.6}{m} length of the detector will need $\sim\SI{1.75}{ms}$ to reach the gas phase. This means that an electron lifetime larger than \SI{1.75}{ms} is required. For the latter, we aim for a gas recirculation flow of at least \SI{70}{slpm} in the purification loop to fully exploit the performance of the xenon purifier. To reach this goal, Xenoscope is built in three phases, with electron drift length increased consecutively: an initial $\sim\SI{50}{cm}$ single phase xenon purity monitor~\cite{Ferella:2006}, followed by a \SI{1}{m} two-phase TPC, and the final \SI{2.6}{m} configuration.

Achieving drift fields of $\sim\SI{200}{V/cm}$ in a TPC with a height of \SI{2.6}{m} requires the delivery of HV ($\sim\SI{50}{kV}$) to the cathode of the detector. The distribution system must be safe and stable, and at the same withstand LXe temperatures. In addition, both its geometry and location in the detector must be optimised to avoid possible distortion of the drift field inside the TPC. Xenoscope was conceived to accommodate different HV systems to identify the optimal configuration for DARWIN, where an extra requirement of radiopurity is needed.

While demonstrating electron drift, the Xenoscope TPC will be instrumented to simultaneously allow for benchmarking of several key parameters relevant for DARWIN. As an example, we mention the measurement of the longitudinal and transverse diffusion of the charge signal at a \SI{2.6}{m} drift length. In a  future campaign, we will investigate optical parameters of LXe, such as the Rayleigh scattering length and index of refraction, via a series of systematic measurements. In addition, Xenoscope will be available to the entire DARWIN collaboration for activities that will range from assessing novel photosensor concepts to testing sub-systems that will be installed in the future DARWIN experiment. 

In this document, we provide an overview of the Xenoscope facility, describe its sub-systems in detail and discuss the performance from a first commissioning run with xenon. The facility infrastructure is described in section~\ref{sec:overview}. The cryostat, its levelling system, the cryogenic and purification systems are introduced in section~\ref{sec:cryo}. We dedicate section~\ref{sec:storage-recovery} to the xenon storage and recovery systems. The core of Xenoscope, the purity monitor and the TPC, are described in section~\ref{sec:tpc} together with their readout systems and the HV design. We introduce the slow control system in section~\ref{sec:SC} and show the first performance data from a commissioning run in section~\ref{sec:commissioning}. Finally, we conclude with a summary of the major achievements and the plans for future measurements in section~\ref{sec:outlook}.

\section{Facility infrastructure}
\label{sec:overview}

\begin{figure}[!ht]
\centering
\includegraphics[width=0.8\textwidth]{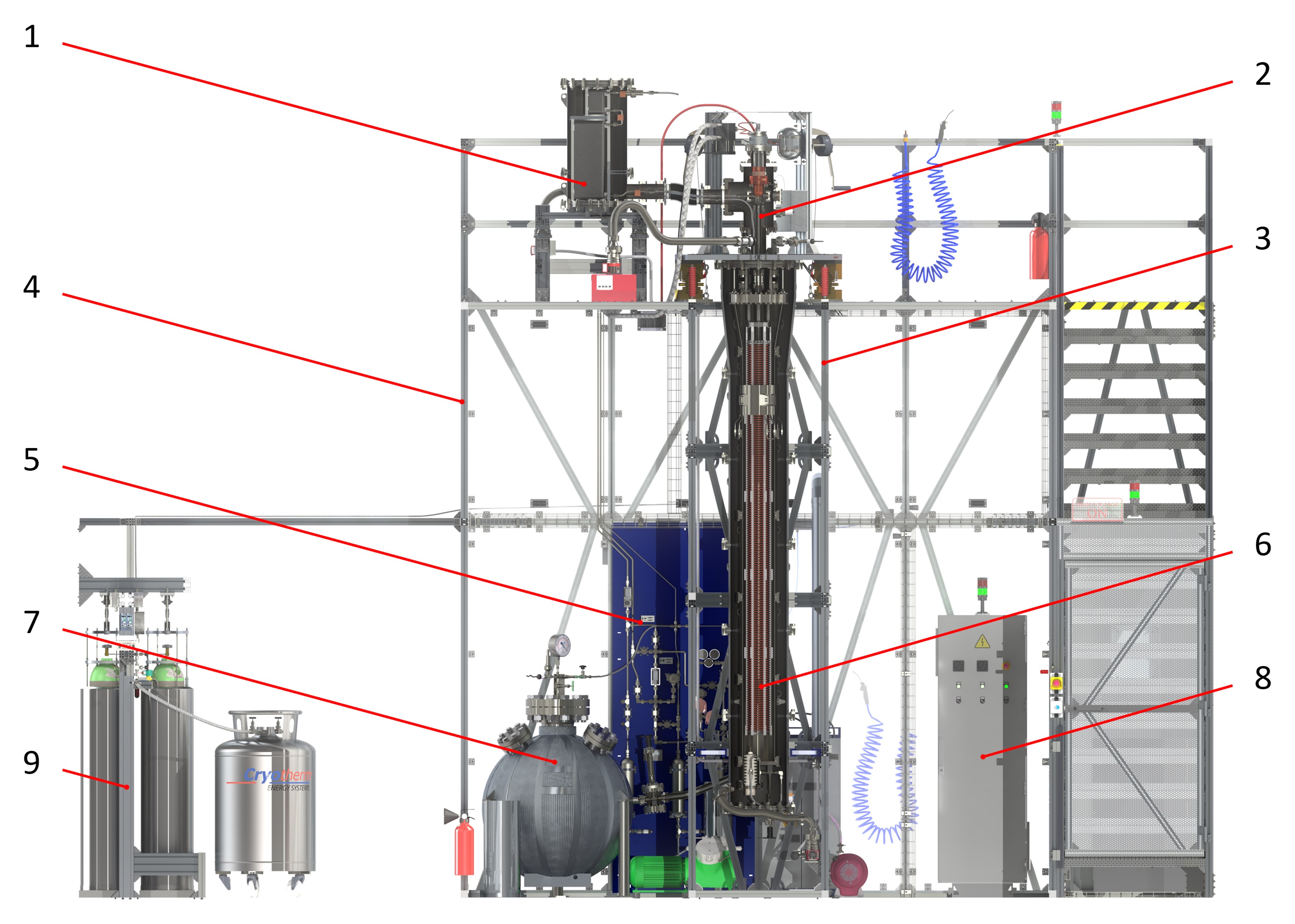}
\caption{Schematic view of the Xenoscope facility with the \SI{2.6}{m} TPC installed in the cryostat. Legend: (1)~Heat exchanger, (2)~Cooling tower, (3)~Inner frame, (4)~Outer frame, (5)~Purification gas panel, (6)~TPC in the \SI{24.8}{cm} diameter and \SI{312}{cm} height cryostat, (7)~High-pressure storage vessel of liquid recovery system, (8)~Power distribution cabinet, (9)~Gas recovery and storage system.}
\label{fig:cross-section}
\end{figure}

Xenoscope, shown schematically in figure~\ref{fig:cross-section}, was designed, constructed and assembled in the high-bay area of the assembly hall of the Department of Physics at the University of Zurich (UZH). The facility is designed to accommodate a full DARWIN-height LXe cryostat. The room is equipped with a 4-tonne bridge crane, a closed-loop water-cooling system, a compressed air system, oxygen level sensors and emergency air exhausts. Two fire-proof storage cabinets are used to store unused gas cylinders. 

\subsection{Support structure}

The support structure is constructed with \SI{50}{mm} $\times$ \SI{50}{mm} extruded aluminium profiles and has three sections: the inner frame, the outer frame and the stairs assembly.  

The inner frame bears the full weight of the \mbox{xenon-filled} cryostat of 1.2~tonnes. It is composed of four stacked segments, each with three vertical weight-bearing profiles. The outer frame with a cubic size of $4 \times 4 \times \SI{4}{m^3}$ provides lateral support to the inner frame and a support for the floor of the upper level. The lower section is enclosed by acrylic panels to minimise the dust accumulation during the assembly of the inner systems. Two double doors grant access to the inside for the cryostat assembly. The upper level can be accessed by a staircase for both assembly and normal operation. The stairs section is fully enclosed by a safety perimeter with interlocks. During operations with high voltage, the interlock secures the access door to the upper level and to the space under the stairs, dedicated to the HV power supply, the data acquisition (DAQ) system, instrumentation rack and slow control server.

Finite element analysis simulations were produced during the design phase as guides to meet all the structural requirements. A final structural study of the support structure was then performed by \textit{Roffler Ingenieure GmbH}~\cite{roffler} to ensure that the structure, in particular the inner frame, could support the estimated \SI{1.2}{tonnes} of the detector and cryostat assembly. In addition, the top platform can withstand an additional load of \SI{200}{kg/m^2}.

\subsection{Electrical power distribution}

The electrical infrastructure of  Xenoscope consists of low- and HV power distribution systems for test voltages up to \SI{100}{kV}.
The electrical scheme was designed to connect all units to a central Earth ground with power supplied by the building mains. An additional distribution path from auxiliary mains is supplied in case of emergency. In addition to the emergency power supplied by UZH generators, a \SI{20}{kW} three-phase Uninterruptible Power Supply (UPS) ensures continuous operation of the system, including all critical equipment such as micro-controllers, servers, vacuum pumps, sensors and the pulse tube refrigerator (PTR).

The Heinzinger PNC 100000-1 power supply allows for voltages up to \SI{100}{kV}(at a maximum current of \SI{1}{mA}) to be applied to the Xenoscope TPC. This unit and the operation of photosensors with voltages $>\SI{1}{kV}$ require additional hardware and procedural measures to ensure safe operation. Thus, a set of redundant safety measures have been implemented, which include: HV shut-down interlocks and keyed doors on areas with restricted access, emergency power-off buttons, \mbox{Lockout-Tagout} procedures to isolate HV equipment, visual alert and alarm systems for when the HV is applied or in the case of failures, and switching command protocols that each user must follow to ensure safe operation. The alert and alarm systems are described in further detail under slow control in section~\ref{sec:SC}.

An electrical safety concept was developed to establish operation, monitoring, and notification protocols, to train users of the facility, and to document the implementation of safety standards in legal compliance with governmental regulations. This document, as well as a full report and electrical schema of the system, were developed together with electrical experts at UZH and approved by an independent, accredited consultant, \textit{Electrosuisse}~\cite{electrosuisse}, in order to ensure adequate implementation of the safety standards.

\section{Cryogenic system and xenon purification}
\label{sec:cryo}

The cryogenic system consists of a double-walled cylindrical stainless steel cryostat, a cooling tower and a heat exchanger. The cryostat is located in the centre of the three weight-bearing profiles of the inner frame and can house the LXe and the various detectors described in section~\ref{sec:tpc}.

To operate the dual-phase Xe TPC, xenon in gaseous form is introduced into the cryostat via a gas handling system and is subsequently liquefied onto a cold head in the cooling tower. The temperature of the cold head is kept constant by a cryo-controller which continuously varies the cooling power of the system. These systems allow for the operation of various experimental setups in both single phase and dual-phase configurations, with continuous gas purification. In this section, we describe the components of the cryogenic and gas handling systems in detail.

\subsection{Cryostat and top flange assembly}
\label{sec:cryostat}

The cryostat is of a modular design to facilitate its assembly. The use of fewer sections reduces the amount of xenon needed for a given test, thus reducing the filling, purification and recuperation times. The outer vacuum vessel is composed of six ISO-K-400 sections. The inner vessel is a pressure vessel capable of containing a maximum of $\SI{400}{kg}$ of LXe, while the final target operational mass is \SI{350}{kg}. It is composed of six electropolished DN250CF tubular sections. Both vessels contain a conical section at their top, extending the diameters to ISO-K-500 and DN350CF, respectively, to make space for the placement of cables and instruments. The usable cylindrical volume of this inner vessel is \SI{24.8}{cm} in diameter and \SI{312}{cm} in height. A full engineering study was performed by \textit{Helbling Technik AG}~\cite{helbling} to ensure the structural integrity of the cryostat both under normal operating conditions at $\SI{2}{bars}$ and $\sim \SI{177}{K}$, as well as in potential emergency situations, in the unlikely event of an uncontrollable increase in pressure. Simulations of the cryostat and CF flanges were performed to confirm their reliability against overload and leakage up to $\SI{8}{bar}$ at $\SI{170}{K}$.

To prevent unwanted movement of the cryostat during normal operation, four locked centering arms constrain the outer vessel at its base. Similarly, adjustable centring arms are installed on the outside of the inner vessel. Contact is made to the inside of the outer vessel via a pointed tip, thus minimising the conductive heat transfer to the inner vessel. The air in the inner vessel can be evacuated after its installation and the vessel can be baked-out to accelerate the outgassing process. The maximum baking temperature must be chosen carefully to not exceed the melting temperature of the various materials used in the setup. Specifically, Multi-Pixel Proportional Counters (MPPCs, see section~\ref{sec:TPC}) should not be heated beyond \SI{60}{\celsius}.

\begin{figure}[!t]
\centering
\begin{minipage}{0.48\textwidth}
\centering
\includegraphics[width=0.8\textwidth]{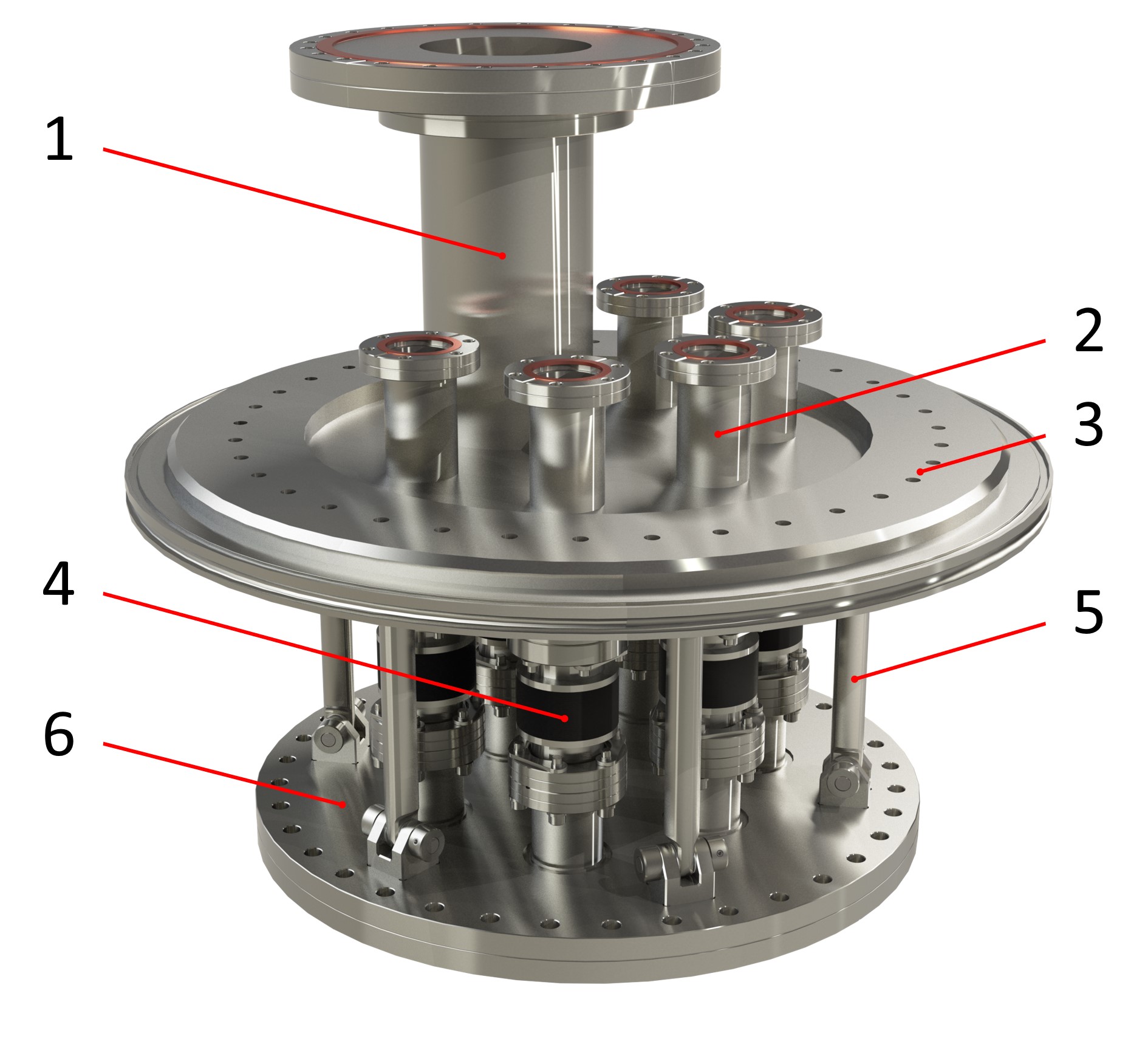}
\end{minipage}\hfill
\quad
\begin{minipage}{0.48\textwidth}
\centering
\includegraphics[width=0.8\textwidth]{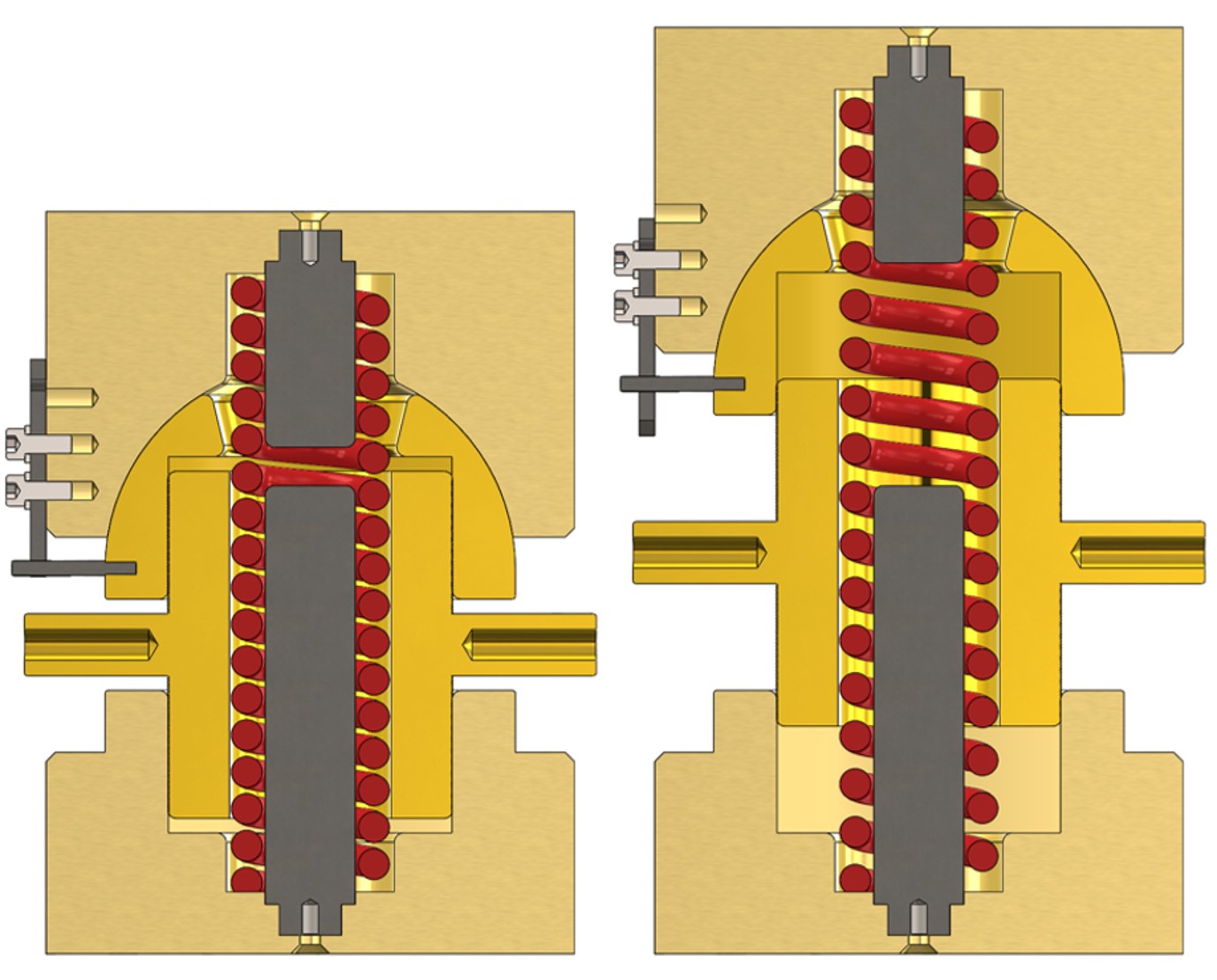}
\end{minipage}
\caption{(Left): Top flange assembly of the cryostat. Legend: (1)~Cooling tower feedthrough, (2)~Instrumentation feedthroughs, (3)~ISO-K500 flange, (4)~Axial displacement bellows, (5)~Swivel rods, (6)~DN350CF flange. (Right): Cross-sectional view one of the levelling legs in fully compressed (left) and fully extended (right) positions.}
\label{fig:top-flange-levelling-leg}
\end{figure}

The top flange assembly of the cryostat, shown in figure~\ref{fig:top-flange-levelling-leg}, left, is composed of an outer ISO-K-500 flange and an inner DN350CF flange. The engineering study showed that a rigid, fully welded design would have been susceptible to welding ruptures caused by the unequal thermal contraction of the two flanges. The bottom half of the top flange, cooled almost to LXe temperature, will contract on the order of \SI{1}{mm} while the top half remains at room temperature. The two flanges are therefore joined by six concentrically-oriented stress-relieving swivel arms of \SI{20}{mm} diameter, suspended by \SI{10}{mm} dowel pins. The six feedthroughs, which host the cooling tower, cabling and instrumentation, are joined by six axial displacement bellows.

\subsection{Cryostat levelling system}
\label{sec:levelling}

To ensure a uniform amplification of the ionisation signal, as well as to prevent large radial forces onto the cryostat and its flanges, the xenon liquid-gas interface must be parallel to the gate and anode meshes. For this purpose, we engineered and commissioned a sophisticated fine adjustment levelling system. To protect this invention, it was registered as a utility model at the German Patent and Trade Mark Office. 

The top flange of the cryostat is bolted to the underside of a $\SI{40}{mm}$ aluminium-magnesium alloy triangular top plate supported symmetrically by three levelling legs that together constitute the levelling assembly. The cryostat is placed in the centre of gravity of the equilateral triangle that the legs form. The legs and the top plate are decoupled by three $\SI{25}{mm}$ thick polyurethane vibration damping plates to reduce the transmission of vibrations onto the detector, which can potentially be produced from instrumentation attached to the support structure. These dampers guarantee an isolation of $\SIrange{50}{90}{\%}$ for excitation frequencies in the range $\SIrange{25}{50}{Hz}$ and $>\SI{90}{\%}$ for frequencies higher than $\SI{50}{Hz}$. 
The core of the levelling assembly are the levelling legs, shown schematically in figure~\ref{fig:top-flange-levelling-leg}, right. Each leg is composed by four major components: a central compression spring, a dual counter-rotating M92 fine thread screw and a hemispherical ball joint that are concentrically aligned, as well as a bottom fixation plate that permits lateral displacement.

The fine thread screw sits in a bottom block on one side and in the hemispherical dome of the ball joint on the other. The dome is mated with a corresponding cavity in the top block. This joint provides a maximum tilting angle of $\SI{5}{\degree}$ between the central axes of the dome and the top block. A steel chain between the bottom and top block ensures that the top block does not lift off the dome. Two small orthogonal pins on the equator, positioned at an angle of $\SI{90}{\degree}$ from each other, are guided by forks on the top block. This ensures that the dome does not rotate inside the cavity while the screw is turned. The spring is compressed between the top and bottom blocks, guided by an outer central tube and inner rods at both ends to prevent it from buckling. The bottom block is firmly bolted to the $\SI{10}{mm}$ thick stainless steel fixation plate, which is itself bolted on top of the inner frame of the support structure. While levelling, the fixation plate allows for a concentric displacement of the legs of $\pm \SI{3}{mm}$. 

The full weight of the filled cryostat rests on the three central springs of the legs, each with a spring constant of $\SI{100.8}{N/mm}$. The counter-rotating fine thread screws with a slope of $\SI{0.75}{mm}$ per turn make for an almost force-free fine adjustment, relieving with ease up to \SI{80}{kg} from each spring. As the major load is carried by the springs, the fine thread stays functional at any time and is easy to turn by hand with short lever arms. Moreover, as the coupling of the bottom and top block is not rigid, and as almost the entire weight rests on the spring, we expect an additional vibration decoupling. The hemispherical joint, together with the concentric degree of freedom given by the fixture plate, allow for a fine horizontal levelling with a precision of $\SI{0.06}{mm/m}$, and accordingly, a precise vertical levelling of the cryostat assembly. The geometry of the legs and the configuration of the springs allow us to level masses up to $\sim \SI{700}{kg}$ per leg. The ball joint and the thread parts are made from brass, a soft metal, which ensures a smooth motion of the rotating and sliding components. As lubricant, a PTFE-based spray is applied.

The operational reliability was demonstrated in a controlled test on a single leg with a mass of $\SI{250}{kg}$ before installation, followed by the successful operation of the full assembly in place on the detector during the levelling procedure of the cryostat. The precision was confirmed with a cross-spirit level with a resolution of $\SI{0.05}{mm/m}$.

The levelling assembly is completed by a three-sided hoisting system, used to assemble the detectors and the vessels of the cryostat underneath it. The hoists are operated in unison by three operators with a spotter who manually guides the lifted section.


\subsection{Pre-cooling system}

\begin{figure}[!t]
\centering
\includegraphics[width=0.6\textwidth]{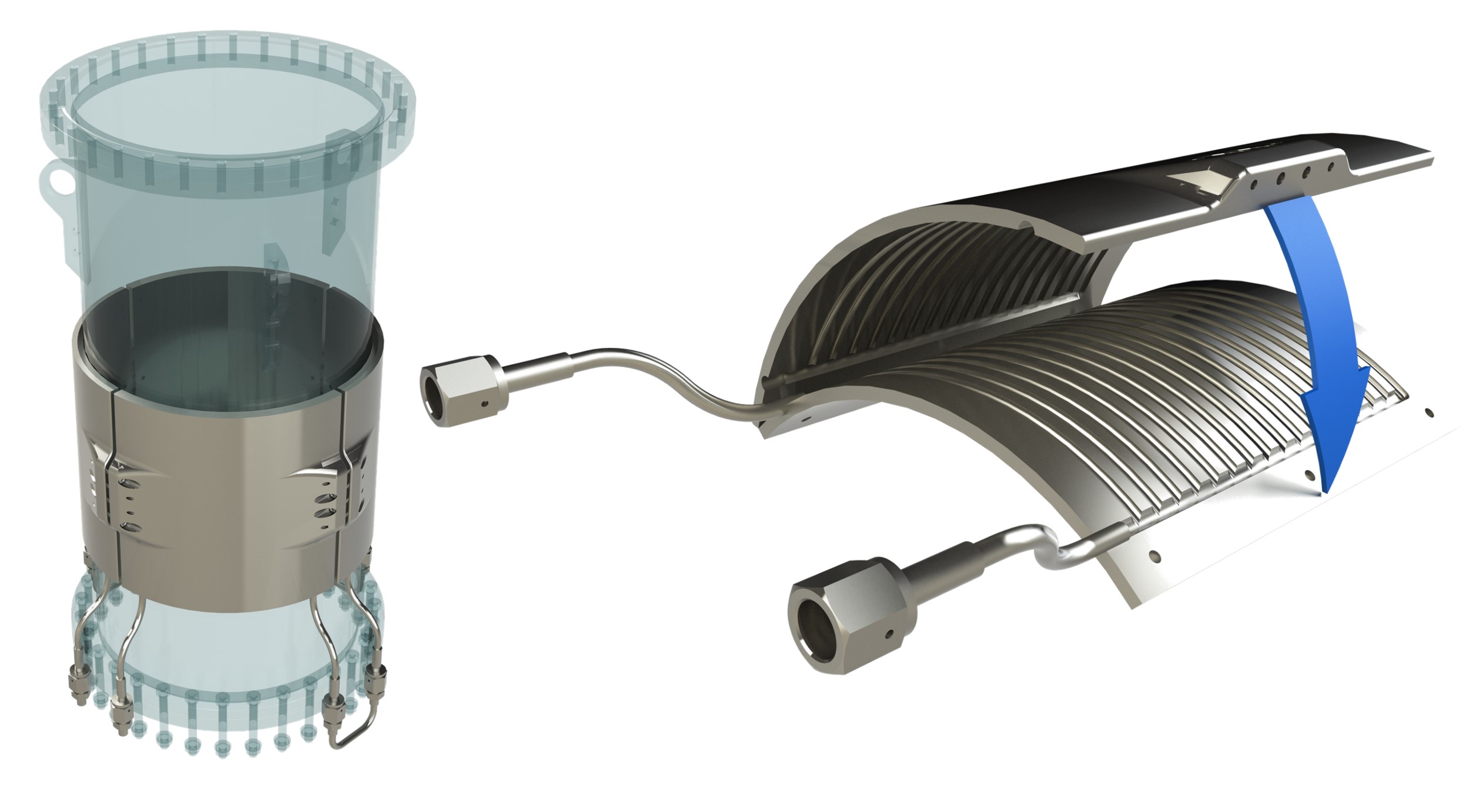}
\caption{Schematic view of the pre-cooler assembly. (Left): Four pre-cooler plates are attached and pressed against the side of an inner vessel section. (Right): Inside view of a pre-cooling plate. Channels are machined inside the stainless-steel plates to distribute the LN$_2$ homogeneously.}
\label{fig:pre-cooler}
\end{figure}

The inner vessel requires an initial cooling period prior to the filling of LXe. To reduce the cool-down time, a pre-cooling system made of four curved stainless steel coolers was built. Shown in figure~\ref{fig:pre-cooler} are four segments connected in series and attached together around the top section of the inner vessel of the cryostat, making direct contact with the cryostat wall. The coolers are made of two machined stainless steel layers, welded at their perimeter. Two full metal hoses connect the coolers to a liquid feedthrough at the bottom of the outer vessel of the cryostat. LN$_2$ is flushed into the coolers from a storage Dewar to pre-cool the inner vessel. The system can achieve a conduction heat transfer of up to \SI{250}{W} at the start of the cool-down process.

\subsection{Cooling tower}
\label{sec:cooling_tower}

\begin{figure}[!ht]
\centering
\includegraphics[width=0.6\textwidth]{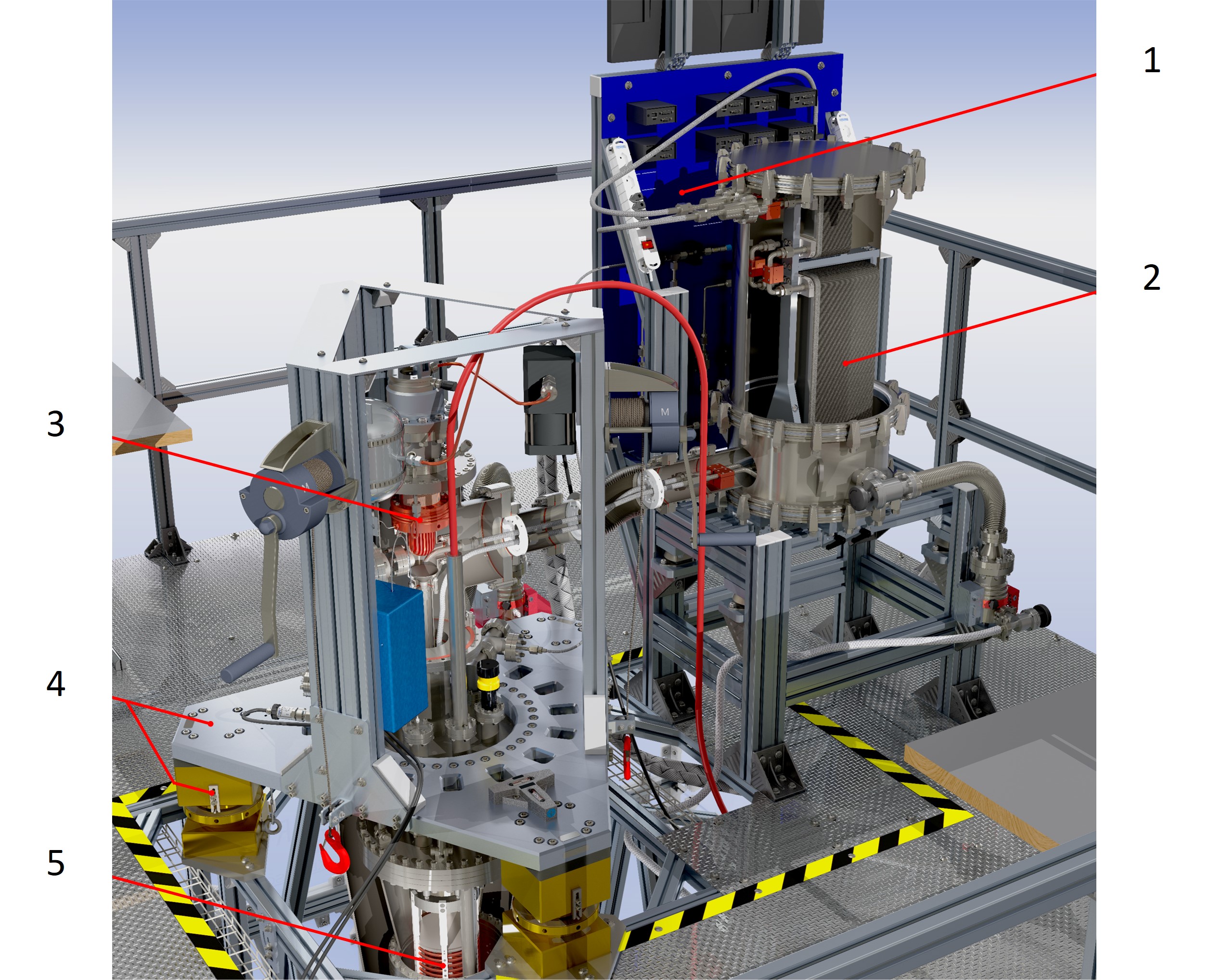}
\caption{Schematic view of the upper floor of the facility. Legend: (1)~Filtration and safety gas system, (2)~ Heat exchanger, (3)~Cooling tower, (4)~Levelling system, (5)~Detector and cryostat.}
\label{fig:top-cross-section}
\end{figure}

\begin{figure}[!ht]
\centering
\begin{minipage}{0.48\textwidth}
\centering
\includegraphics[height=6.7cm]{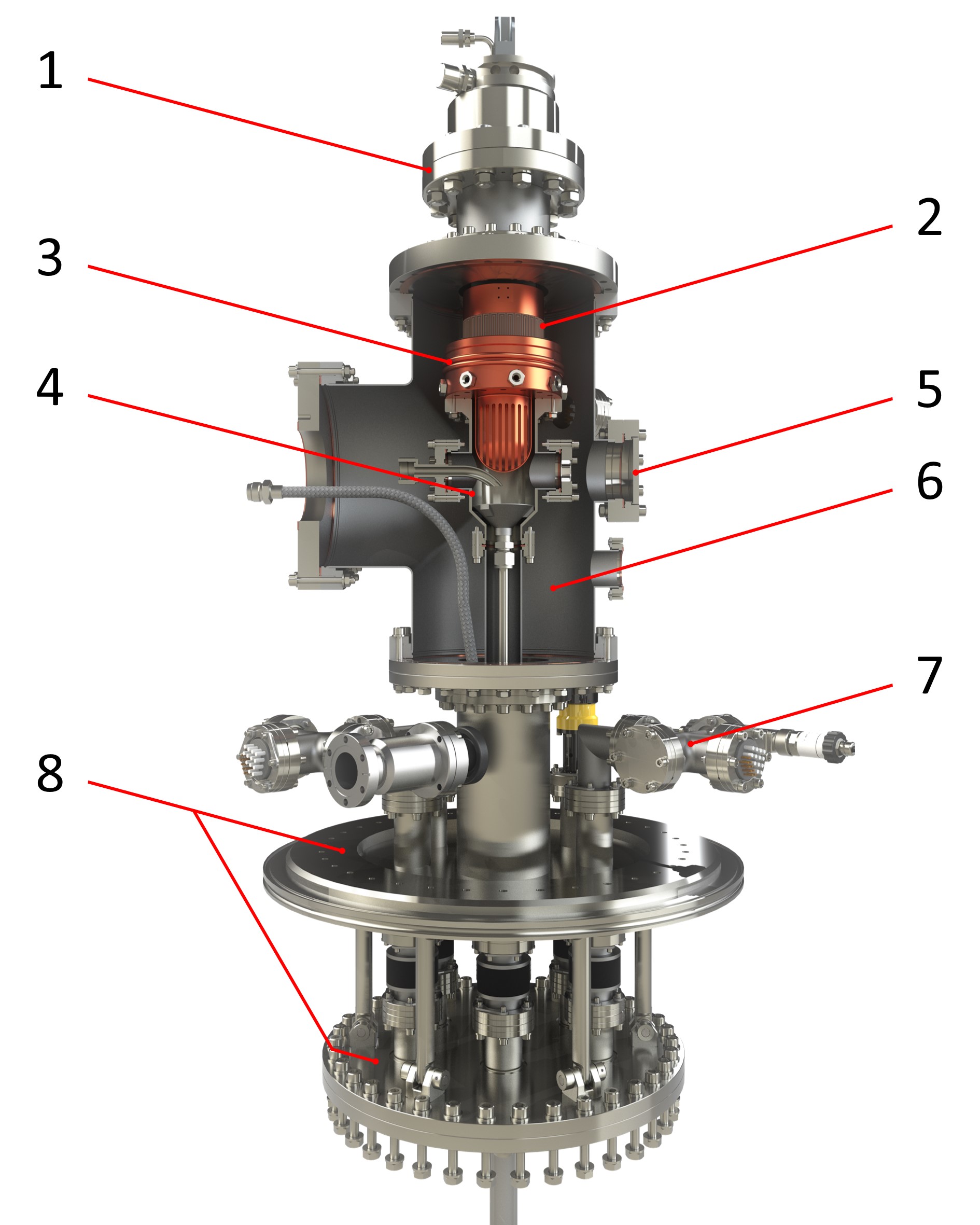}
\end{minipage}\hfill
\quad
\begin{minipage}{0.48\textwidth}
\centering
\includegraphics[height=6.7cm]{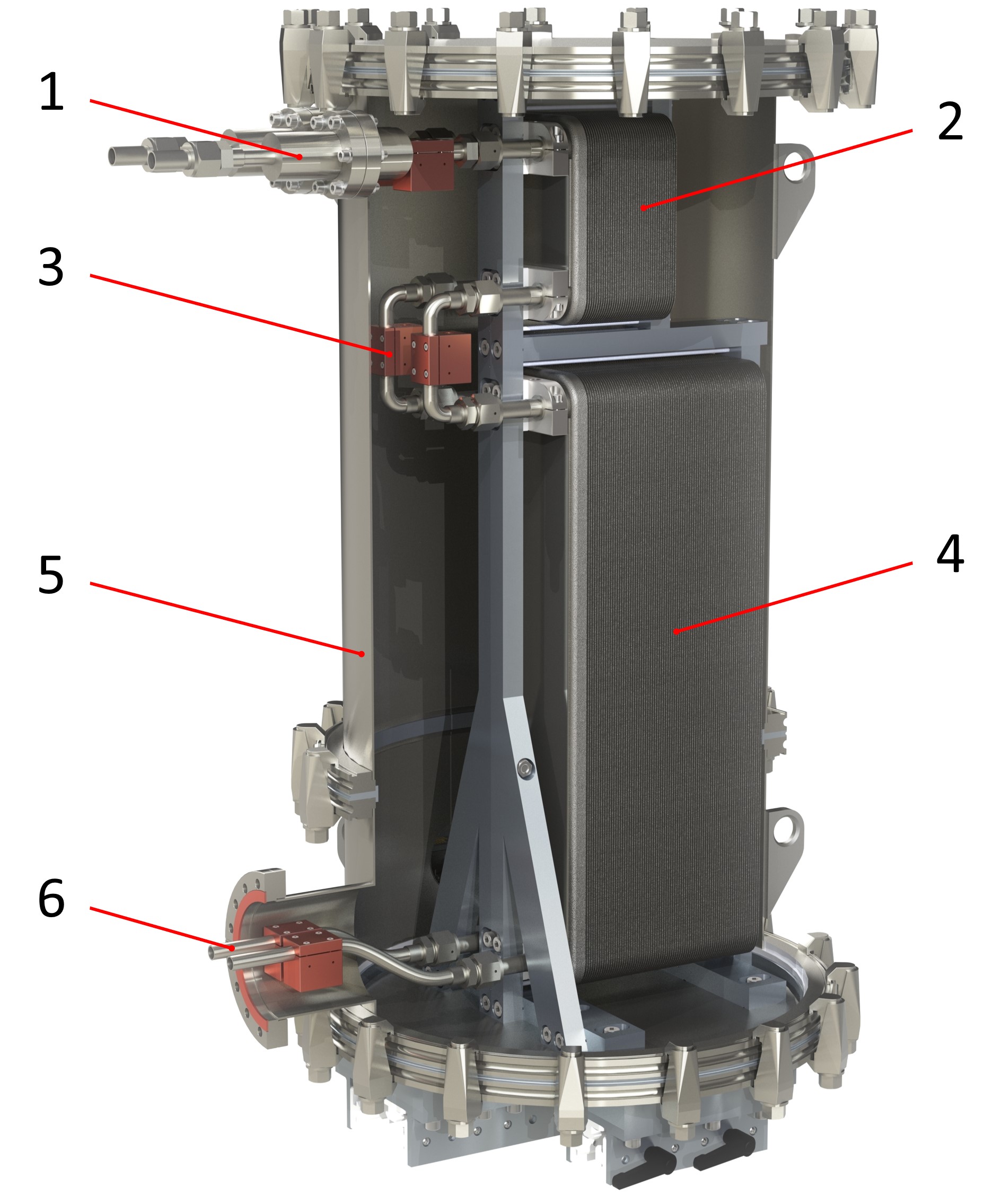}
\end{minipage}
\caption{(Left): Schematic view of the cooling tower installed on the top-flange. Legend: (1)~PTR, (2)~Heater, (3)~Cold head, (4)~Cooling chamber, (5)~Viewport, (6)~Vacuum chamber, (7)~Feedthroughs ($6\times $), (8)~Top flange. (Right): Schematic view of the heat exchanger system in its vacuum chamber. Legend: (1)~Gas inlet/outlet, (2)~Pre-cooler/post-heater, (3)~Temperature sensors ($6\times$), (4)~Vaporiser/condenser, (5)~Vacuum chamber, (6)~Liquid inlet/outlet.}
\label{fig:CT-HE}
\end{figure}

The cryogenic system is responsible for the liquefaction and the overall thermal stability of the facility. Two main sections can be identified: the cooling tower and the heat exchanger system. The cooling tower, placed on top of the facility and shown in figures~\ref{fig:top-cross-section} and~\ref{fig:CT-HE}, left, provides the necessary cooling power to condense xenon and maintain it in liquid form. At the top is an \textit{Iwatani Corporation} PTR~\cite{iwatani}, model PC-150, rated to $\sim \SI{200}{W}$ at $\SI{165}{K}$. A $\SI{180}{W}$ heater unit allows for a precise temperature control of the system. The PTR cools down the cold head, onto which the gaseous xenon (GXe) condenses, drips down into a funnel and is transported to the bottom of the cryostat. Four PT100 Resistance Temperature Detectors (RTDs) are screwed into the side of the cold head for temperature monitoring and control. During normal operation, the temperature is kept between \SIrange{173}{175}{K} to obtain a constant pressure of $\sim \SI{2.0}{bar}$.

A secondary LN$_2$ cooling system is also integrated into the cooling tower design. A coiled copper tube is affixed to the outside of the cold head, providing cooling by flushing LN$_2$ through the tube. Cryogenic thermal paste is used to maximise heat transfer. The oxygen-free copper cold head is installed on top of a stainless-steel cooling chamber, using an indium-sealed flange. Two aligned viewports are mounted on the side of the cooling chamber for visual observations of the liquefaction process.

The cooling chamber and other stainless steel parts have been cleaned in ultrasonic baths with acidic soap and distilled water for at least \SI{20}{min}, followed by a rinse in the ultrasonic bath in pure ethanol. The copper cold head and mating plate were cleaned with the same method, with an additional ultrasonic bath in a solution of citric acid at \SI{2}{wt\%}, followed by a final ethanol rinse.

\subsection{Heat exchanger system}

In spite of the recent development of LXe recirculation and purification systems, the removal of impurities from the gas phase is still required. The growth of LXe detectors to multi-tonne scale and the resulting requirement to recirculate the GXe at high speeds have made the use of heat exchangers indispensable~\cite{Giboni:2011wx,Aprile:2012jh,Akerib:2012ya}. Heat exchangers with high efficiencies minimise the heat leak from cold outgoing and warm incoming GXe and thus allow the purification of high amounts of xenon in the gaseous phase in short time periods. This is necessary as the cooling power of commercially available PTRs is limited (see section~\ref{sec:cooling_tower}). Connected to the cooling tower via a flexible umbilical section is a cascade of two brazed stainless-steel plate heat exchangers, produced by \textit{Kelvion Holding GmbH}~\cite{kelvion}. The two heat exchangers, seen in figure~\ref{fig:CT-HE}, right, have respectively a heat exchange area of $\SI{4.6}{m^2}$ (Model GVH 500H-80) and $\SI{0.66}{m^2}$ (Model GVH 300H-40). They are installed in series inside a vacuum vessel and allow for high recirculation rates through the purification loop with minimal heat loss in the cryostat. To ease assembly and adapt to movements of the levelling assembly, the position of the vacuum chamber is adjustable both in height and distance, relative to the cooling tower. In addition, the entire heat exchanger cryostat rests on dampening mounts to reduce the transmission of possible vibrations onto the heat exchangers and the sensitive equipment inside the vacuum chamber. To minimise heat conduction over the inner holding structure, the heat exchangers are decoupled by plastic blocks at the attachment points on the fittings and on the back. Before installation, both heat exchangers were flushed with water and soap and dried with ethanol. They were then baked out for over a week at temperatures up to \SI{180}{\celsius} in order to efficiently remove liquids and outgassing vapours.

Recirculation through the purification loop is performed by removing xenon from the top of the liquid phase with the underpressure (relative to the cryostat pressure) created by the xenon compressor in the return line. The reduced pressure induces a phase-change from liquid to gas and lowers the temperature of the latter by expansion. This creates a temperature gradient with the counter-flow in the supply line which is necessary for an efficient heat exchange process and crucial for phase changes that naturally happen at zero temperature difference~\cite{Aprile:2012jh}. The xenon is sent through the vacuum-insulated umbilical connection to the heat exchanger system. It passes through the larger vaporiser where, depending on the recirculation flow, some xenon could accumulate in liquid form, and then through the smaller \mbox{post-heater}, which produces the remaining gas-gas heat exchange. Concurrently, on its way back from the purification loop in the supply line, the xenon is pre-cooled in the smaller heat exchanger and condensed in the larger one. As we shall see in section~\ref{sec:commissioning_he_efficiency}, the condensation actually happens inside the umbilical connection for the whole tested flow range. The purified xenon enters the cooling tower and is reintroduced at the very bottom of the inner vessel. Extraction from the top and re-introduction at the bottom induces a natural upward flow in the cryostat, ensuring that all the xenon is purified. This configuration has a calculated heat exchange power of $\SI{1.5}{kW}$ for the described process. For performance monitoring, a total of six PT100 RTDs are installed in copper housings, in thermal contact with the heat exchanger connections, as shown in figure~\ref{fig:CT-HE}.

\subsection{Thermal insulation}

Every cold section of the facility (cryostat, cooling chamber and heat exchanger systems) is thermally insulated against heat convection and conduction with vacuum and \mbox{multi-layer} insulation (MLI) to limit radiative losses. The vacuum system is composed of two turbo-molecular pumps from \textit{Pfeiffer Vacuum GmbH}~\cite{pfeiffer} located at the bottom of the outer vessel of the cryostat and at the vacuum chamber of the heat exchanger. Both turbo-molecular pumps are evacuating the same volume and are backed up by one common backing pump. 

The MLI was designed and produced by \textit{RUAG Space GmbH}~\cite{ruag}. It is composed of three blankets of 10 layers of $\SI{12}{\micro m}$ thick polyester sheets, aluminised at \SI{400}{\angstrom} on both sides. Because of space constraints, only two blankets are used on the heat exchanger. The layers are perforated for use in vacuum with an open area of $\SIrange{0.05}{0.1}{\%}$. Each layer is separated by a layer of non-woven polyester spacer at $\SIrange{14}{15}{g/m^2}$. Since the cryostat will be opened several times during the lifetime of the facility, the superinsulation of the cryostat is assembled with stitched Velcro strips. The rest of the MLI is installed with cryogenic tape, making it a permanent installation. 

\subsection{Gas handling and purification system}
\label{Sec:Gas_handling_and_purification}

\begin{figure}[t!]
\centering
\includegraphics[width=0.9\textwidth,origin=c]{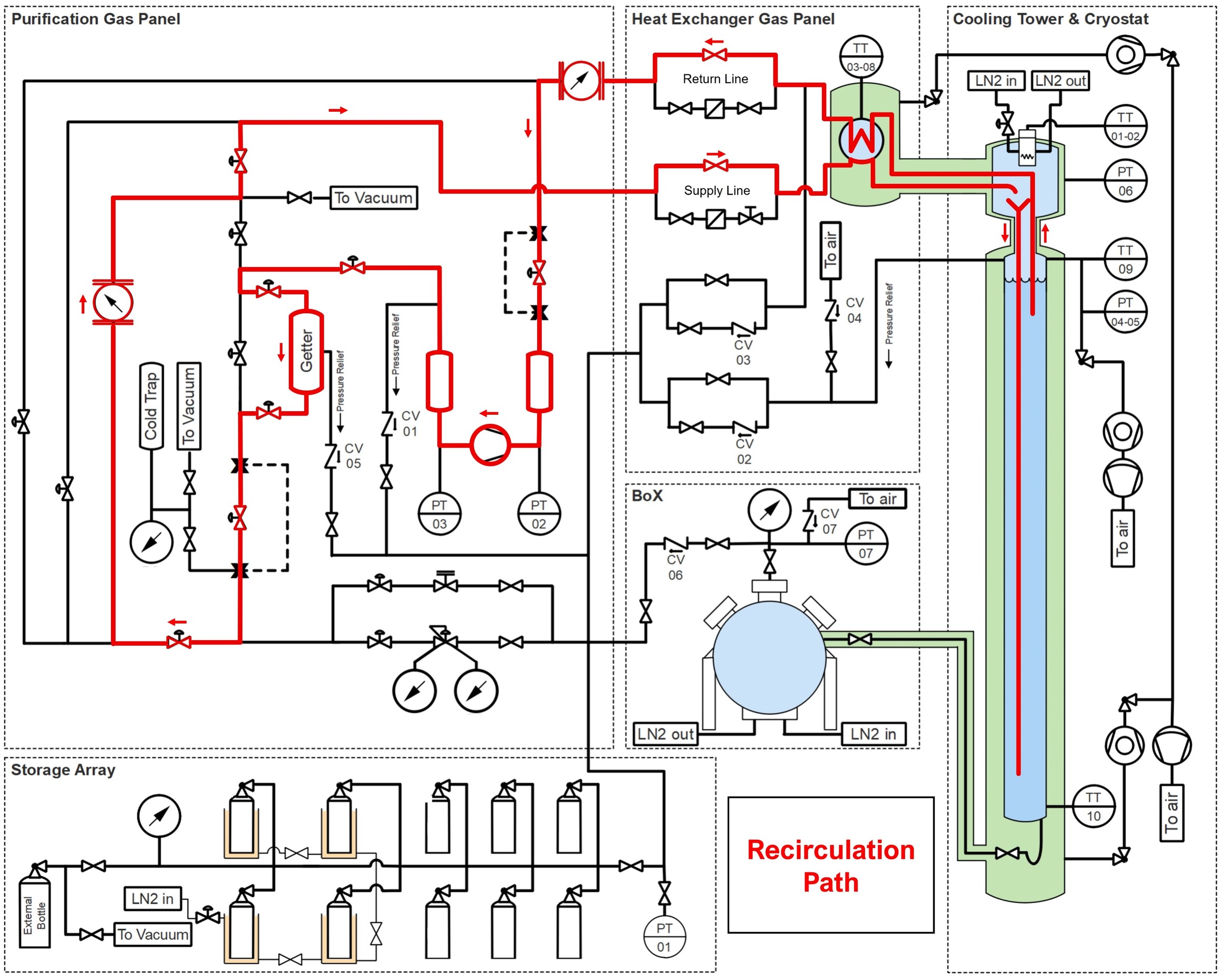}
\caption{Piping and Instrumentation Diagram of the Xenoscope facility. The path in red shows the xenon purification loop. LXe is taken through the heat exchanger, whereby heat is absorbed from incoming GXe to vaporise it. The GXe flow is controlled by a flow controller, restricting the passage of xenon upstream of the compressor. GXe is passed through the hot metal getter, and is sent back to the heat exchanger where it is condensed to LXe before returning to the cryostat through the cooling tower.}
\label{fig:PID}
\end{figure}

The gas handling system is used for the filling of GXe from the storage systems to the cryostat, for the purification of the GXe and for its recuperation back into the storage systems. Figure~\ref{fig:PID} shows the Piping and Instrumentation Diagram (P\&ID) of the entire system. The main gas panel hosts several components of the purification loop: the xenon purifier, a PS4-MT50-R-535 hot metal getter manufactured by \textit{SAES}~\cite{saes}, factory certified for flows of up to $\SI{70}{slpm}$ of xenon and capable of handling over $\SI{100}{slpm}$; a double-diaphragm xenon compressor, model N1400.1.2SP.12E by \textit{KNF Neuberger}~\cite{knf}, which produces a pressure difference of up to \SI{3}{bar}, achieving a theoretical maximum flow of up to $\SI{250}{slpm}$. The flow of GXe is controlled by a HFC-203 flow controller produced by \textit{Teledyne Hastings}~\cite{teledyne}, which is located before the xenon compressor. In addition, a HFC-201 flow meter is located downstream of the compressor.
 
To minimise the pressure drop due to turbulence, \SI{0.5}{inch} pipes, fittings and valves are used in the purification loop. The main gas panel also features dual-acting pneumatic valves, which can be actuated remotely. This provides an additional level of safety by reducing reaction time in case of an emergency, allowing for the remote activation or deactivation of the purification loop. The valves are actuated by a valve controller, supplied with air from the air-supply system of the building at \SI{7}{bar}. A nitrogen bottle regulated at \SI{6}{bar} serves as a backup in case of the unlikely failure of the main air supply system. Two lift check valves provide an uninterrupted supply in air, the nitrogen bottle being used when the pressure of the main line drops below \SI{6}{bar}.

Two buffer volumes are installed inline with the compressor to protect it from sudden pressure spikes. The main gas panel also features a cold trap, which can be used to achieve an ultra-high vacuum in the whole system prior to the initial filling. The cold trap can also be used to cryo-pump and freeze the GXe present in the gas lines in order to safely perform maintenance without the need for a full recuperation.

Complementing the main gas system, the heat exchanger unit features a secondary gas panel for the emergency recuperation line and particulate filters. On the xenon extraction line, a $\SI{40}{\micro m}$ metal strainer element protects the purification loop from particulates. On the xenon supply line, a ceramic ultra-high purity filter removes particulates of $\SI{3}{nm}$ size with an efficiency of nearly \SI{100}{\%}.

In case of an unexpected pressure increase anywhere in the gas system, the emergency line directs xenon released by pressure-relief valves to the gas storage and recovery system, where it is collected in an LN$_{2}$-cooled aluminium bottle (see section~\ref{Sec:Gas_recovery}). The pressure relief valves are located at strategic locations in the system: CV-01 protects the xenon compressor by relieving pressure from $\SI{4.0}{bar}$ absolute; CV-02 serves as the first pressure protection of the cryostat volume by releasing gas at pressures greater than $\SI{3.0}{bar}$; CV-03 is located at the exit of the heat-exchanger system, limiting the pressure at the exit to $\SI{4.0}{bar}$; CV-05 protects the getter system from pressures greater than \SI{11.3}{bar}. Finally, as a last resort, in case of an uncontrollable pressure rise above the \SI{8.0}{bar} absolute cryostat pressure limit, CV-04 can vent the GXe to air. The flow through all filters and check valves can be shut off by in-line manual valves. Furthermore, all filters and check valves, except for CV-04, can be bypassed. 
\section{Xenon storage and recovery systems}
\label{sec:storage-recovery}

For safe long-term storage of the xenon during maintenance, upgrade work and downtime, the gas handling system connects to dedicated recuperation systems. We distinguish a baseline system for the recuperation in the gas phase and a system that recuperates the xenon in the liquid phase to accelerate the recuperation procedure significantly. Both are designed and built.

\subsection{Gas recovery system}
\label{Sec:Gas_recovery}
\begin{figure}[h!]
\centering
\begin{minipage}{0.42\textwidth}
\centering
\includegraphics[width=1\textwidth]{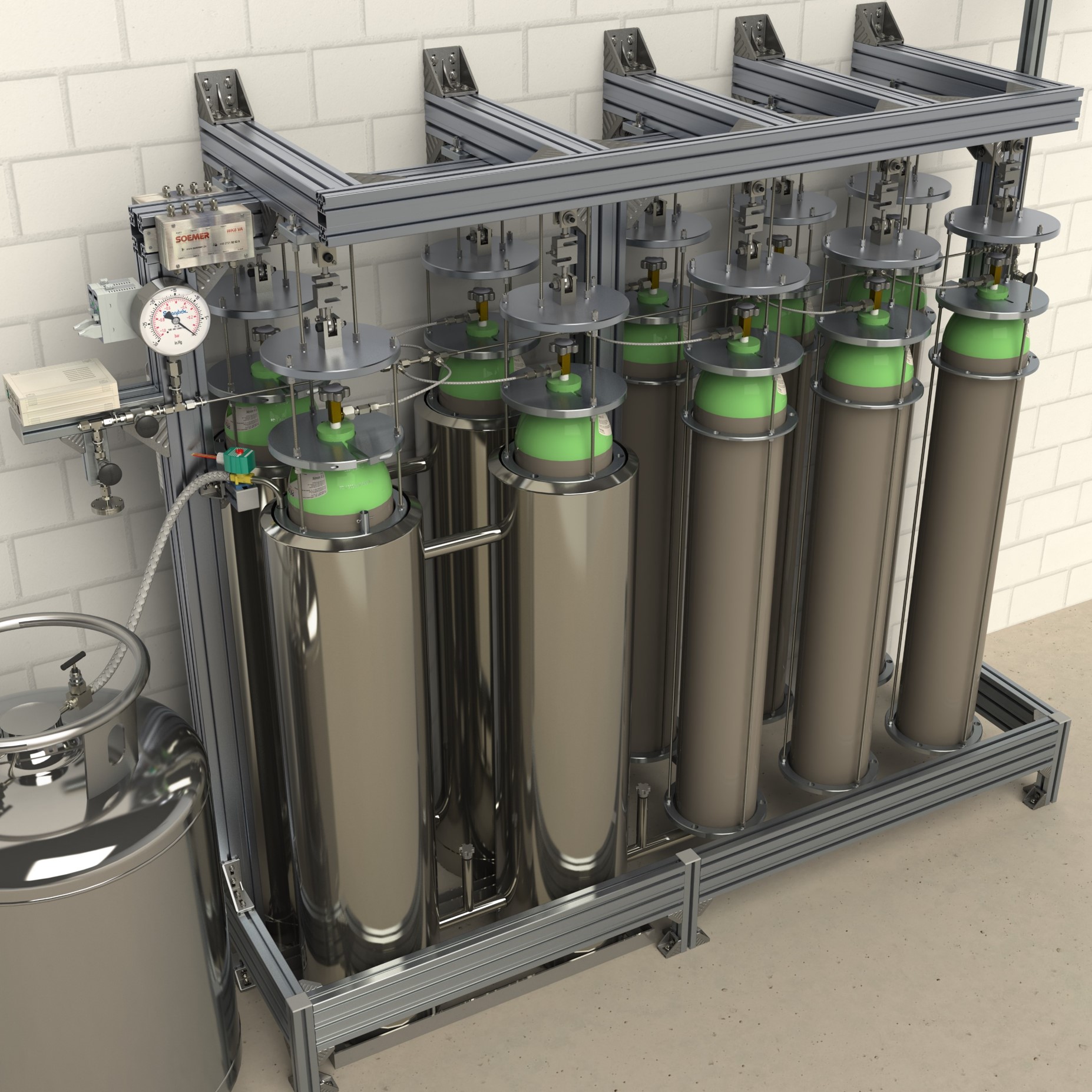} 
\end{minipage}\hfill
\quad
\begin{minipage}{0.52\textwidth}
\centering
\includegraphics[width=0.95\textwidth]{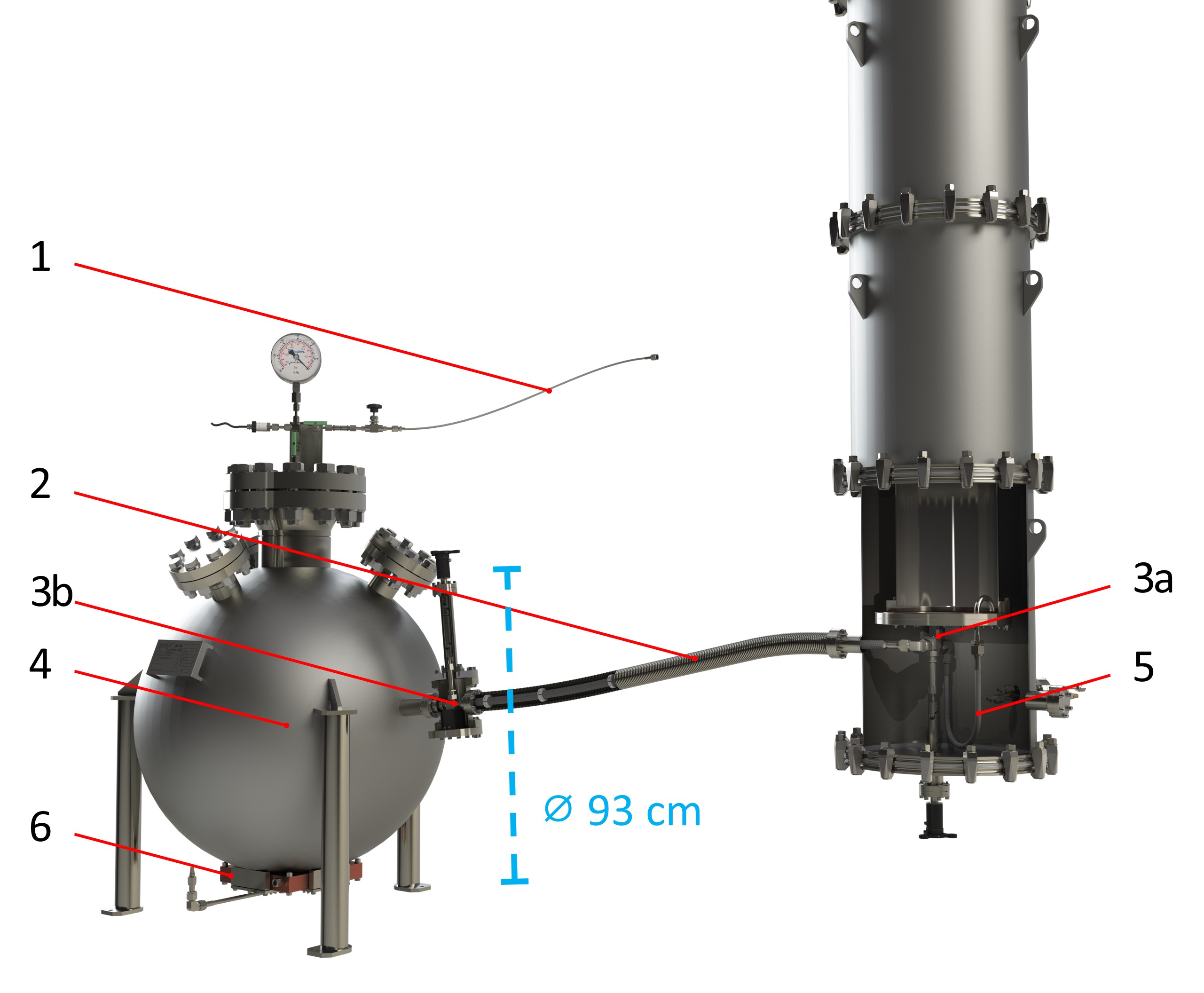}
\end{minipage}
\caption[Recuperation and Storage Systems]{(Left): Gas recuperation and storage system. Ten aluminium gas cylinders are suspended from load cells for mass determination, four of which are placed inside an inter-connected Dewar flask assembly. (Right): Schematic view of the liquid recuperation and storage system. Legend: (1)~Gas return line to high-pressure side of the gas system, (2)~Vacuum-insulated liquid transfer line, (3)~Cryogenic valves, (4)~Spherical pressure vessel BoX, (5)~Cryostat drain with cane-shaped inlet and P-trap, (6)~LN$_2$ cooler. For better visualisation, the TPC and the HV feedthrough are hidden.}
\label{fig:recup_sys}
\end{figure}

As shown in figures~\ref{fig:PID}~and~\ref{fig:recup_sys}, left, a gas bottle storage array consisting of ten $\SI{40}{l}$ aluminium gas cylinders is connected to the gas handling system on the high-pressure side, allowing us to store a maximum of $\SI{470}{kg}$ of xenon. Four of the gas cylinders are suspended in tall, inter-connected open Dewar flasks from \textit{Cryofab, Inc}.~\cite{cryofab} (Model CF1248) that are $\SI{122}{cm}$ deep at full diameter. The Dewars feature bottom drains via which they are connected in series by means of vacuum insulated lines. Individual Dewars can be isolated from the others via vacuum-jacketed valves. Cane-shaped inlets and P-traps protect the valves from the accumulation of moisture. Depending on the xenon mass to be recuperated, a set number of Dewars is filled with LN$_{2}$. The xenon is then evaporated from the cryostat and recovered into the cold cylinders by means of cryogenic pumping. The four cold cylinders can hold the full xenon amount in liquid and solid phase. To reduce condensation and for safe operation of the bottle valves, band heaters with a power of $\SI{1400}{W}$ each, located at the top of each bottle, keep the valve and the neck above the freezing temperature of water. 

After recuperation, the xenon is equalised among all ten cylinders in gas phase during warm-up. The recuperation line features an analog pressure gauge with $\SI{250}{bar}$ full-range and a vacuum gauge with an admissible pressure of $\SI{10}{bar}$ which can be isolated by a valve. All components exposed to high-pressure are rated to $\SI{200}{bar}$ or higher.  The first cylinder in the array is connected via the recuperation and emergency line to the check valves (see section~\ref{Sec:Gas_handling_and_purification}). As part of the passive safety system, this bottle is kept cold at any time during operation, resulting in a pressure of $\SI{\sim e-1}{mbar}$ in the line. In the event of an automatic recuperation through the check valves, the vacuum gauge can be used to detect the increased pressure in the recuperation line. 

A levelling system from \textit{KGW-Isotherm}~\cite{KGW} controls the LN$_{2}$ level of the first Dewar. A total of six PT100 sensors are installed, four for two 2-point level controls at different heights and two for under- and overfill alarming. A bottle weighing system allows for the monitoring of the xenon inventory and indicates any xenon loss during run cycles. The suspended gas cylinders are supported by \textit{Mettler-Toledo Schweiz GmbH}~\cite{Mettler-Toledo} (Model SLS510) $\SI{250}{kg}$ capacity load cells that feature an individual combined error of $\leq \SI{45}{g}$. Two \textit{Soemer Messtechnik GmbH}~\cite{soemer} (Model WK8 VA) junction boxes sum up the analog signals of five load cells each. The signals are then read out by two \textit{Mettler-Toledo} (Model ACT350) terminals.

\subsection{Liquid recovery system}
\label{sec:BoX}
Tests of multiple new technologies with Xenoscope inevitably involve frequent xenon filling and recuperation processes. A fast and safe recovery system with the potential for long-term storage is therefore essential. While the speed of gas recovery is limited by the heat input onto the inner vessel of the cryostat required to vaporise the LXe, transfer of xenon in liquid form is faster at lower LN$_2$ consumption. To this end, a separate LXe recovery system was designed and constructed. At its heart is a cryogenic pressure vessel into which LXe is drained by gravity~\cite{Virone:2018zjy} from the bottom of the cryostat, which we call BoX (Ball of Xenon). BoX can store up to $\SI{450}{kg}$ of xenon in both liquid and gas phase.
 
BoX is installed close to the cryostat, the liquid inlet below the bottom outlet of the cryostat (see~figure~\ref{fig:recup_sys}, right). A vacuum insulated $1/2"$~cryogenic line connects the inlet of BoX on its side to the bottom of the cryostat. At its top, a warm $1/4"$~gas line connects the outlet of BoX through the high-pressure side of the gas system, to the cold head of the cooling chamber. When filled with LXe, the hydrostatic pressure of the xenon column leads to an increased total pressure of $\sim \SI{3}{bar}$ at the bottom of the fully-assembled inner vessel of the cryostat. Located in the insulation vacuum and actuated by a rotary feedthrough, a cryogenic bellows sealed valve with secondary containment system and a spherical metal stem tip (Swagelok~\cite{swagelok} model SS-8UW-V47-TF) seals the inner vessel. A P-trap and the fact that the bottom drain is located at the cryostat wall protect the detector volume from gas bubbles that are possibly created at the warmer, potentially gaseous, dead end. Depending on the amount of xenon filled in the detector, BoX will be empty or at a certain gas pressure during detector operation. Its liquid inlet is sealed by a second vacuum-jacketed cryogenic valve of the same type. Its gas outlet features another valve of that kind in the $1/4"$~version.

Engineering provided by the manufacturer \textit{KASAG Swiss AG}~\cite{kasag} led to a final draft of the pressure vessel based on our conceptual design. To ensure its safety, \textit{KASAG Swiss AG} performed detailed calculations regarding the rules and standards AD~2000 according to PED 2014/68/EU for pressure devices~\cite{PED}.
The stainless steel pressure vessel was chosen to be spherical to minimise its mass and thus, its heat capacity, which is key for the cryogenic design to reduce nitrogen consumption during the initial cool-down. In addition, a low mass reduces the cost of materials and facilitates handling during installation. The inner radius $r$, which directly influences the required wall thickness for a given amount of xenon contained at an overpressure $p$, was optimised for mass based on:  
\begin{equation*}
m(r) = \mathrm{FOS} \cdot \frac{4}{3}\pi r^3 \left[ \left( 1+\frac{p}{\sigma} \right)^{3/2} -1 \right] \rho,
\end{equation*}
where FOS denotes the required factor of safety on the yield strength $\sigma$ of the stainless steel, while $\rho$ is its density. The equation follows from the equilibrium of the tensile strength of the cross-section of a hollow sphere and the force  from the inner overpressure acting onto the cross-section. For a hollow sphere, $m(r)$ is constant for large $r$ and divergent for small $r$. Unlike for ideal gases, $m(r)$ additionally features a minimum at moderate $r$ for a Van der Waals gas. For a capacity of $\SI{450}{kg}$ of xenon, we obtain the lowest mass at an inner radius of $r= \SI{450}{mm}$.

The vessel is designed for a maximum working pressure of $\SI{90}{bar}$ and for temperatures down to $\SI{-196}{\celsius}$. It is equipped with a proportional safety relief valve (CV-07) set to the maximum working pressure, and it is instrumented with an analog pressure gauge and a pressure transducer for pressure monitoring. The inner surface, including the welds, is pickled and electropolished through the service ports at the top to improve the inside cleanliness. Every gasket in contact with xenon is entirely made of metal. The sphere is held by three welded legs bolted to the floor. Since the maximum liquid level is located $\sim \SI{70}{mm}$ below the equator, the legs are welded at the side, above the equator, to reduce heat conduction. Underneath the sphere, a soldered LN$_2$ copper cooler with an internal spiral channel is mounted such that it is held and pressed against the surface by four welded brackets. For good thermal contact, a mixture of cryogenic Apiezon N grease and $\SIrange{0.5}{1.0}{\micro m}$ silver powder is applied in between the sphere's surface and the cooler.

For xenon recovery, BoX is pre-cooled, which liquefies remnant xenon inside. The cooling power is pressure-controlled for this procedure to keep the pressure between $\SIrange{2}{3}{bar}$ inside BoX. When the valves on the liquid line between BoX and the inner vessel of the cryostat are opened, the pressure in BoX rises to $\sim \SI{3}{bar}$ and the connecting line starts cooling from LXe evaporation. By opening the valves on the gas return line, GXe is brought back to the cryostat, replacing the LXe flowing into BoX. The speed of the recovery process can be controlled either with the valves on the liquid and gas line or with the pressure regulator on the main gas panel. A check valve (CV-06) on the gas line avoids back-flow of gas in case of pressure fluctuations. 

BoX is single-walled and passively insulated with multiple layers of aluminised bubble wrap that features a reflective shield to prevent radiative heat absorption. The cooler region is additionally insulated by a milled foam part. During recuperation, we expect a need of $\sim \SI{200}{W}$ cooling power due to radiation, conduction and convection processes. This corresponds to an LN$_2$ consumption of $\sim \SI{3.6}{kg/h}$. At low or zero recirculation flows, the spare power of the PTR on the cryostat will be sufficient to cover most of this demand. At the end of the recovery process, the remaining GXe in the system can be cryo-pumped into the gas bottles or into BoX by freezing the xenon.

While isochorically warming up, the LXe will change to gas. Above its critical pressure of $\SI{58.4}{bar}$, xenon is in a supercritical state~\cite{NIST}. At room temperature, the pressure in BoX is $\sim \SI{72}{bar}$ when loaded with $\SI{450}{kg}$ xenon. In its supercritical state, the pressure of xenon is highly temperature dependent. However, below $\SI{40}{\celsius}$, the pressure will not exceed the working pressure of the vessel and hence, the set pressure of the safety relief valve.

The detector is filled with GXe through the pressure regulator on the main gas panel. Xenon condenses at the cold head cooled by the PTR, as it is the case when filling from the gas bottles.
\section{Purity monitor and time projection chamber}
\label{sec:tpc}

\begin{figure}[ht!]
\centering
\includegraphics[width=0.6\textwidth]{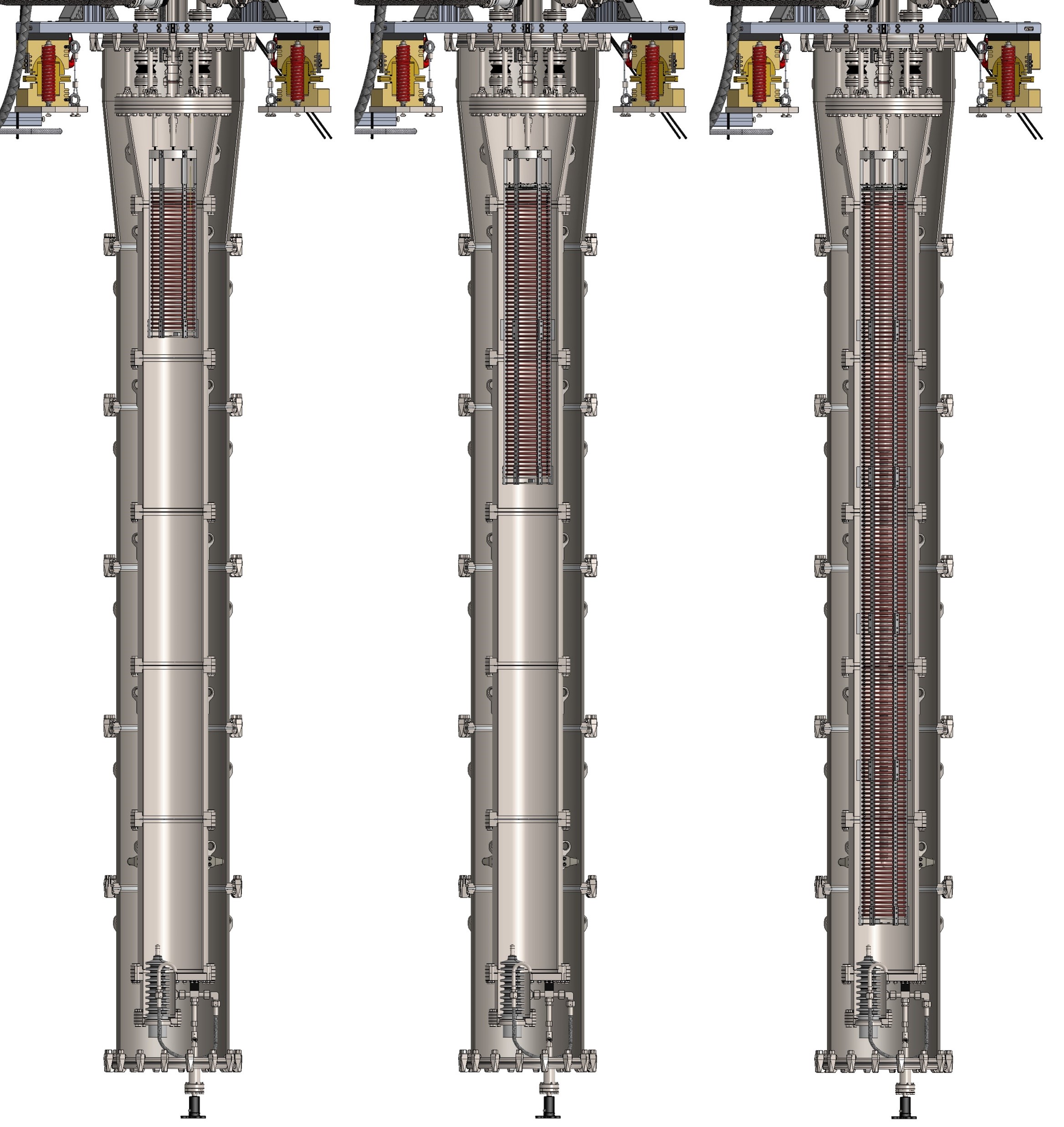}
\caption{The three phases of Xenoscope. From left to right: purity monitor, \SI{1.0}{m} tall TPC, \SI{2.6}{m} tall TPC. All three phases are built from the same modular sections of the drift cage. The purity monitor uses two charge readouts, while an MPPC array replaces the top charge readout in the \SI{1.0}{m} and \SI{2.6}{m} TPCs.}
\label{fig:phases}
\end{figure}

The measurement campaign to demonstrate the \SI{2.6}{m} drift will take place in three stages. In the first phase, electrons will be drifted in a \SI{525}{mm} long Purity Monitor (PM) fully immersed in LXe, where the trigger is provided by the flash of a xenon lamp. This will provide an initial measurement of the achievable electron lifetime. In the second stage, the PM will be upgraded to a dual-phase \SI{1}{m} tall TPC with liquid level control and a signal amplification region will be added to the system. This will allow for the extraction of electrons from the liquid into the gaseous phase, thus generating electroluminscence. An array of MPPCs will be placed at the top to detect the resulting proportional light signal. In the final \SI{2.6}{m} dual-phase TPC configuration, a HV ceramic feedthrough will be installed, as described below, allowing for an applied cathode voltage of \SI{-50}{kV} in order to achieve the required drift field for the \SI{2.6}{m} electron drift length. The three detectors are shown in the cryostat in figure \ref{fig:phases}.

\subsection{Purity monitor} 
\label{sec:PM}
Impurities with high electronegativity in the LXe can capture drifting electrons on their path, reducing signal strength. Common impurities in commercially available LXe consist of $\sim$ppm levels of O$_{2}$, N$_{2}$, H$_{2}$O, as well as organic molecules~\cite{Hasterok:2017ehi}. Other detector and subsystem materials also introduce these impurities by outgassing. An O$_2$ concentration of less than \SI{1}{ppb} is necessary to successfully drift electrons along the \SI{2.6}{m} TPC~\cite{Bakale:1976,Aprile:2009dv}; such a purity is obtained by constant purification of the GXe through a hot metal getter which traps electronegative impurities. The impurity concentration in LXe can be estimated by monitoring the electron lifetime inside the TPC, via drifting of electrons in the detector volume and measuring the charge loss due to electronegative impurities after a given distance. In the PM, free electrons are produced by pulses of high-intensity photons generated from a flash lamp, incident on a photocathode. PMs have been implemented in experiments such as ICARUS~\cite{Amoruso:2004ti}, ProtoDUNE~\cite{Diurba:2020rmm} and XENONnT~\cite{Moriyama:2019}. 

\begin{figure}[ht!]
\centering
\includegraphics[width=1\textwidth]{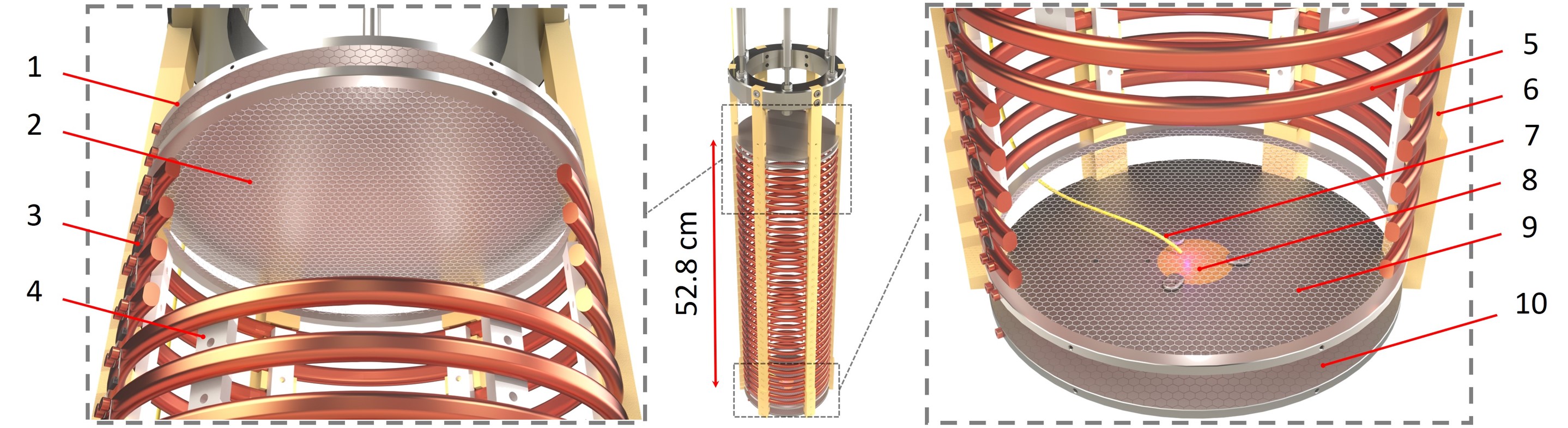}
\caption{Schematic view of the PM, consisting of four parallel electrodes (1, 2, 9, 10). The photocathode (10) is flashed by a xenon lamp via an optical fibre (7) pointing to centre of the photocathode (8) which is coated with a thin metal film. The copper field-shaping rings (5) are held by polyamide-imide pillars (6) and locking blocks (4). The resistor chain (3) provides the potential drop to shape the electric field.}
\label{fig:purity_monitor}
\end{figure}

The design of the PM for Xenoscope is shown in figure~\ref{fig:purity_monitor}. It consists of four parallel electrodes, where the photocathode is flashed by a xenon lamp via an optical fibre pointing at its centre.
The size of the pulse coming from the electrodes depends on the number of photons incident on the photocathode. The population of photoelectrons induces a signal in the photocathode as they drift towards the first screening mesh positioned \SI{17}{mm} above, producing a maximum once it enters the second drift region. These electrons are drifted towards the top by the applied electric field, across a distance of \SI{525}{mm}. When the electron cloud reaches the third drift region, a second signal is acquired at the anode as the electrons move away from the grid, which is positioned \SI{10}{mm} below, until they are fully collected. The flash of the lamp provides a clear trigger for the data acquisition.

The ratio between the two charge signals is proportional to the electron lifetime and yields an estimation of the electronegative impurity concentration. The electron arrival time can be used together with electric field simulations, discussed in section~\ref{sec:HV}, to estimate the uniformity of the strength of the electric field along the drift path, as a change in drift velocity relative to the predicted value can indicate field inhomogeneity.

The field cage is composed of modular sections of \SI{525}{mm} which can be connected together to produce the \SI{1}{m} and \SI{2.6}{m} tall drift regions. It consists of oxygen-free high conductivity copper rings of 15 cm inner diameter, a total of 173 for the \SI{2.6}{m} TPC, joined by interlocking PTFE pieces affixed to the inner side of the field cage. The copper rings are suspended by 6 pillars made of polyamide-imide, selected due to its good mechanical, electrical and low out-gassing properties. The pillars are connected to a stainless steel ring, suspended under the top flange by six stainless steel rods. The HV in the PM phase is supplied by a NIM power supply through a \SI{10}{kV} \textit{CeramTec} SHV feedthrough.

The radiometric properties of the xenon lamp in the spectral region below the work function of the selected photocathode material, the attenuation of UV photons in the light guide, the quantum efficiency (QE) of the photocathode material, and the light loss from the coupling between the different systems are parameters that affect the expected photoelectron production.

The xenon lamp is a \textit{Hamamatsu} L7685 flash lamp \cite{hamamatsu}, with \SI{1}{J} of input power. The incident spectrum ranges from 200 to \SI{800}{\nm}, with $\sim\SI{10}{\%}$ of the emission above the work function of the photocathode materials, below \SI{6}{eV}. The photocathode consists of a quartz substrate and a thin metal film, custom-made using an accelerated argon ion sputter with an integrated quartz micro-balance. The ion sputter allows us to test different coating materials and to control the thickness of the thin film (on the order of \SI{0.5}{nm}). We are currently testing gold, silver and aluminium as potential candidates. The thickness of the thin film is optimised depending on the electron escape probability and the absorption length of the material used. The quartz substrate reflects the non absorbed light, enhancing the release of photoelectrons in the thin layer. The optical fibre is a $\SI{600}{\micro m}$ diameter polyimide coated silica fibre, resistant to photodarkening and suitable for low temperatures. The PM construction and commissioning follows after an initial test of the electronics and the photocathode. Various photocathode materials are benchmarked in a smaller setup in vacuum, GXe and LXe, with the charge readout electronics.

\begin{figure}[ht!]
\centering
\includegraphics[width=1\textwidth]{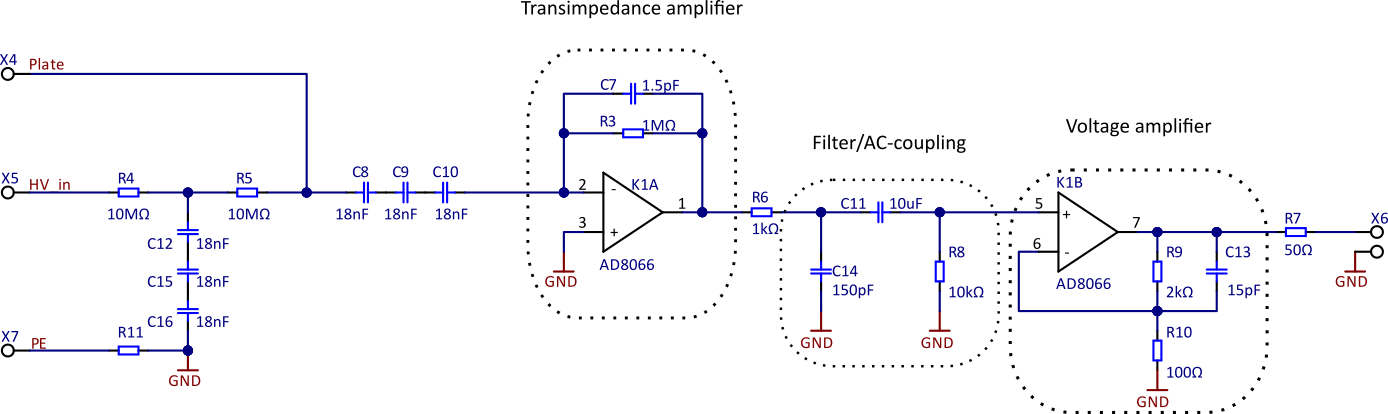}
\caption{Circuit design for the cathode signal readout in the PM, with HV stabilising filter and a two-stage signal amplification.}
\label{fig:circuit}
\end{figure}

The readout electronics consists of a circuit board with two distinctive functional parts: one HV filter to stabilise possible current variations, and two instances of amplification once the signal is decoupled from the HV component. The components of the board must fulfill two fundamental functions: ensure a safe and stable delivery of HV to the photocathode and provide low-noise electronics for the amplification of small signals.
Figure~\ref{fig:circuit} shows a schematic of the circuit.

\begin{figure}[ht!]
\centering
\begin{minipage}[t]{0.48\textwidth}
\centering
\includegraphics[width=1\textwidth]{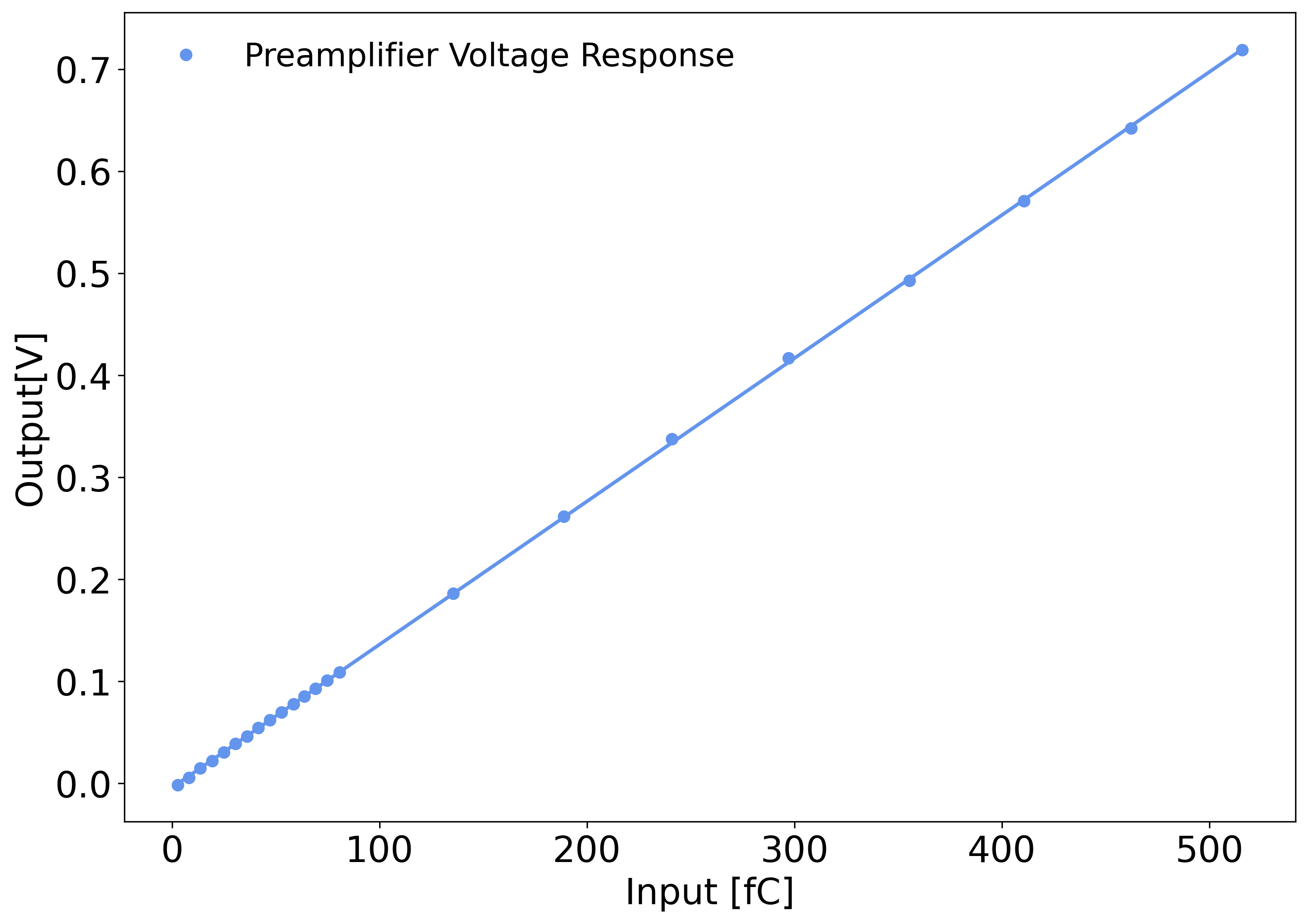}
\end{minipage}\hfill
\quad
\begin{minipage}[t]{0.48\textwidth}
\centering
\includegraphics[width=1\textwidth]{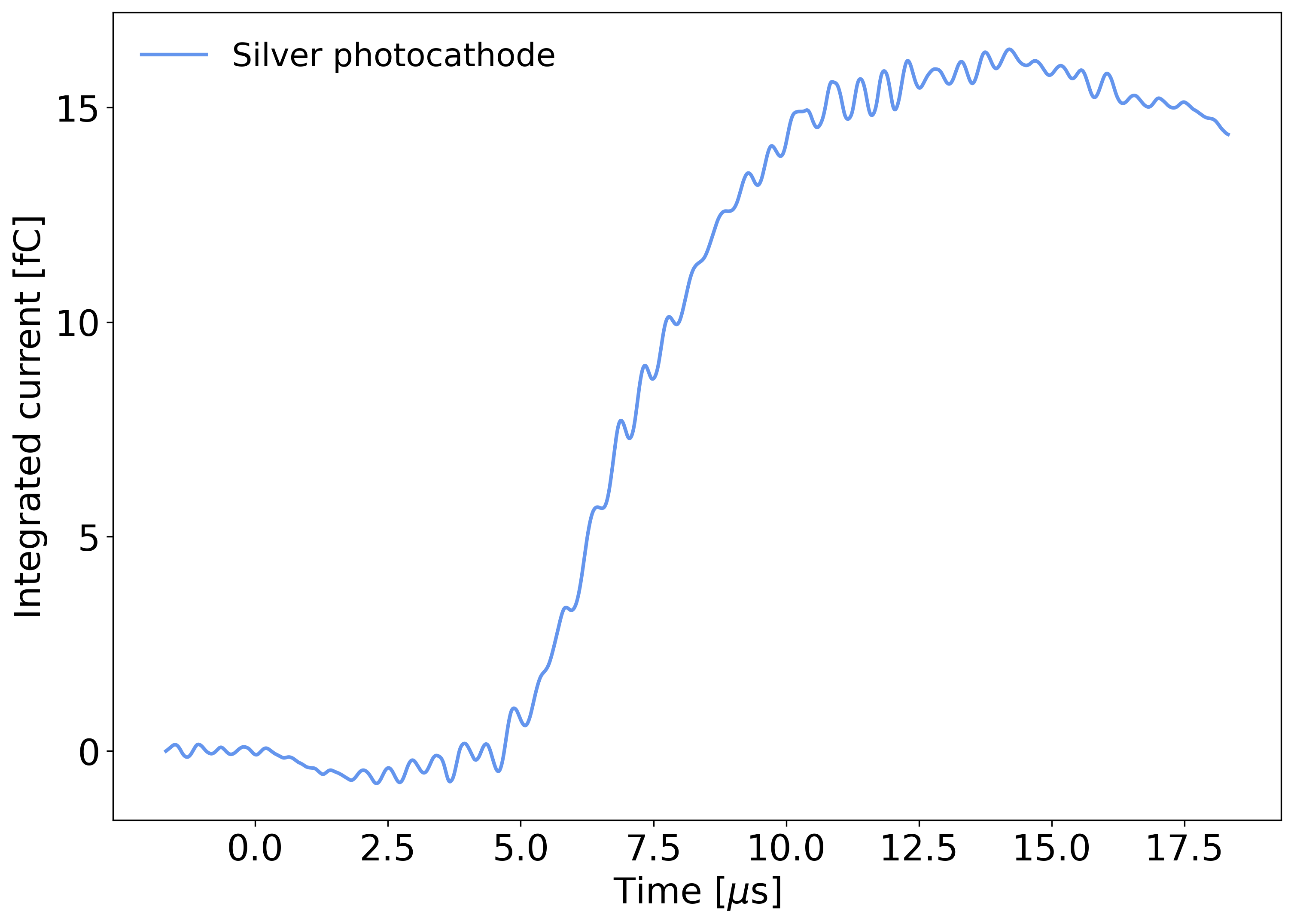}
\end{minipage}
\caption{(Left): Characterisation of the charge preamplifier board response to input charge. The response is linear in the range of the estimated charge that electrons travelling through the grids will induce. (Right): Integrated current seen at the output of the preamplifier circuit in a tri-electrodic configuration in LXe with a silver photocathode. }
\label{fig:liquid_xenon_silver}
\end{figure}

The charge readout circuit is placed in LXe next to the cathode to reduce the impact of thermal noise, of the leakage current and of the mutual capacitance between the detector and the meters-long cables, on the amplified signal. The first tests with the board involved the measurement of the response of the circuit at low temperatures, the withstanding of the applied HV, the measurement of the response in frequency and the measurement of the noise level. To this end, an environmental chamber and a LXe setup described in~\cite{Barrow:2016doe} are used, in which the signals are acquired with an oscilloscope and the trigger is provided by the flash of the xenon lamp.  We show the response of the circuit at $\SI{200}{K}$ in the temperature-controlled environmental chamber in figure~\ref{fig:liquid_xenon_silver}, left, where the overall good linearity in the range of tens to hundreds of femtocoulomb can be observed. The linear fit yields an amplification value of \SI{1.4}{V/pC}. Figure~\ref{fig:liquid_xenon_silver}, right, shows the signal acquired in LXe at \SI{176}{K} with a silver photocathode, in a tri-electrodic configuration with a hexagonal etched pattern as screening mesh. The electron collection in this configuration was not optimised and the shown results are only a characterisation of the electronics response. The duration of the pulse is dominated by the input discharge signal from the xenon lamp. The preliminary test also resulted in an upgrade of the enclosure of the xenon flash lamp to further reduce electromagnetic interference originating from each discharge. 

\subsection{Dual-phase time projection chamber}
\label{sec:TPC}

After the first electron lifetime measurements with the PM, the inner detector will be replaced by a dual-phase TPC. The design of the $\SI{1}{m}$ TPC and the final $\SI{2.6}{m}$ detector are identical apart from the replacement of the \SI{10}{kV} SHV feedthrough with a bottom mounted \SI{100}{kV} ceramic HV feedthrough, as described in section~\ref{sec:HV}, as well as the addition of field cage modules for the latter. In the TPC operational mode, the primary signals will be generated in the form of electrons via the PM photocathode, as previously described, rather than from xenon scintillation induced by particle interactions (traditionally called the ``S1'' signal). This choice was motivated by the \mbox{$\sim16{:}1$} height-to-width aspect ratio of the TPC, which severely limits the light collection efficiency throughout the bulk of the detector as shown by Monte Carlo simulations. However, using a photocathode to generate the primary signal in a TPC has several advantages. Data acquisition will be triggered by the flash lamp, thus providing an unambiguous charge signal generated at the photocathode. This feature, in combination with the known drift distance between the fixed photocathode position and the liquid/gas interface, will facilitate pairing of the primary (charge) signal with its correlated secondary signal detected at the top of the TPC. Event reconstruction may otherwise be a challenge in such a system, given the expected high background rate due to operation above ground without shielding. As opposed to a continuous acquisition by means of a discriminator module, this pulsed acquisition mode also minimises the amount of data collected. Furthermore, using electrons as the primary signal eliminates the need for PTFE reflectors which are known to largely contribute to the outgassing rate in a xenon TPC~\cite{Bruenner:2020arp}, thus reducing the load on the gas purification system.

\begin{figure}[ht!]
\centering
\includegraphics[width=1\textwidth]{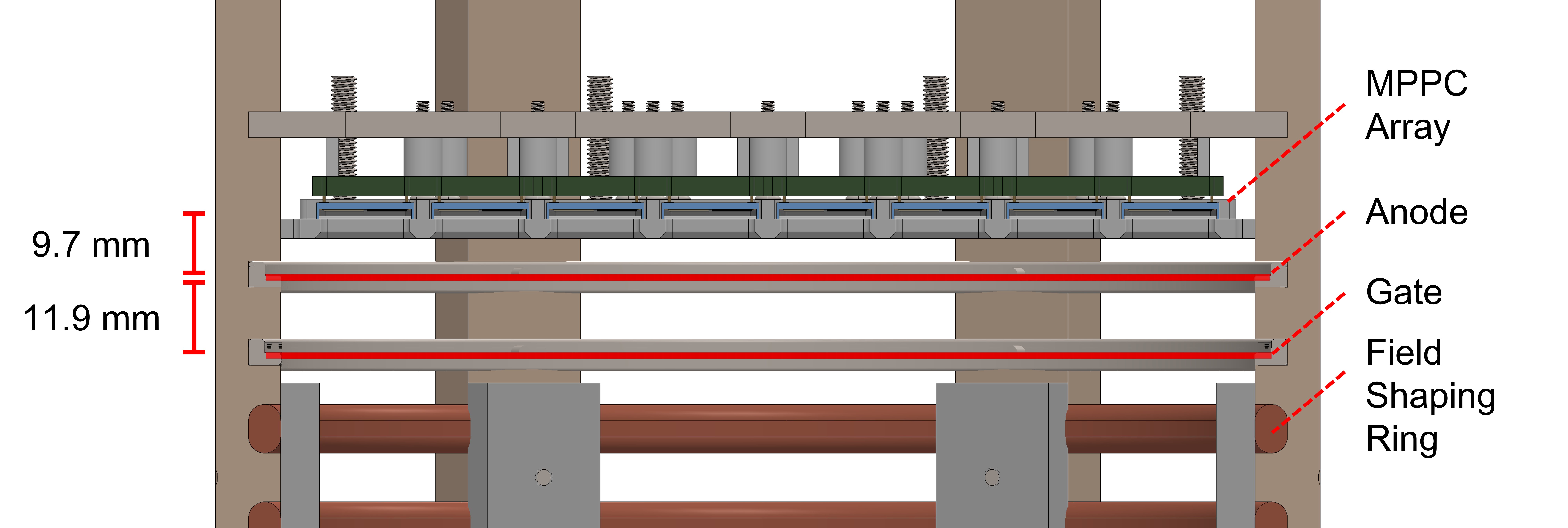}
\caption{Geometry of the top section of the TPC. The two \SI{0.1}{mm} thick} meshes (gate and anode) are highlighted in red and displayed ten times thicker for clarity. The GXe/LXe interface is located between the gate and the anode. The height of the interface is determined empirically by optimising the light yield of the ionisation signal.
\label{fig:TPC-top}
\end{figure}

In the TPC mode, a liquid/gas interface will be implemented and the charge collector will be replaced with an anode mesh in the gaseous phase. The drifting electrons will be extracted from the LXe to the GXe phase with a nominal electric field of up to \SI{10}{kV/cm}, thus producing a secondary (scintillation) signal via electroluminescence at the liquid/gas interface. The geometry of the top section of the TPC is shown in figure \ref{fig:TPC-top}. The electrode meshes are made from \SI{0.1}{mm} thick, chemically etched stainless steel with \SI{93}{\%} transparency. The gate and the anode are separated by 11.9\,mm while the MPPC array is located 9.7\,mm above the anode.

The amplification factor of the secondary signals in a dual-phase TPC depends on the strength of the extraction field and also on the distance of the liquid level to the anode. The liquid level is controlled with a weir system, using a magnetically coupled linear motion feedthrough. This design allows for the level to be adjusted by up to \SI{8.4}{mm}. The weir consists of an open-top stainless steel cylinder and a center-channel, made of two tube sections and an edge-welded bellows, welded concentrically. The centre-channel is filled with LXe, which can flow out of a hole pierced in the top tube. Compression or expansion of the bellows with the linear motion feedthrough adjusts the height of this hole with a precision of $\SI{2.5}{\micro m}$, which modifies the LXe height. The weir is mounted onto a stainless steel mounting tube using PTFE clamps which extend down from the top flange. For the future operation of a top-down HV feedthrough, the stainless steel mounting tube will be removed, and the weir will be directly clamped onto the HV rod.

\begin{figure}[ht!]
\centering
\includegraphics[width=1\textwidth]{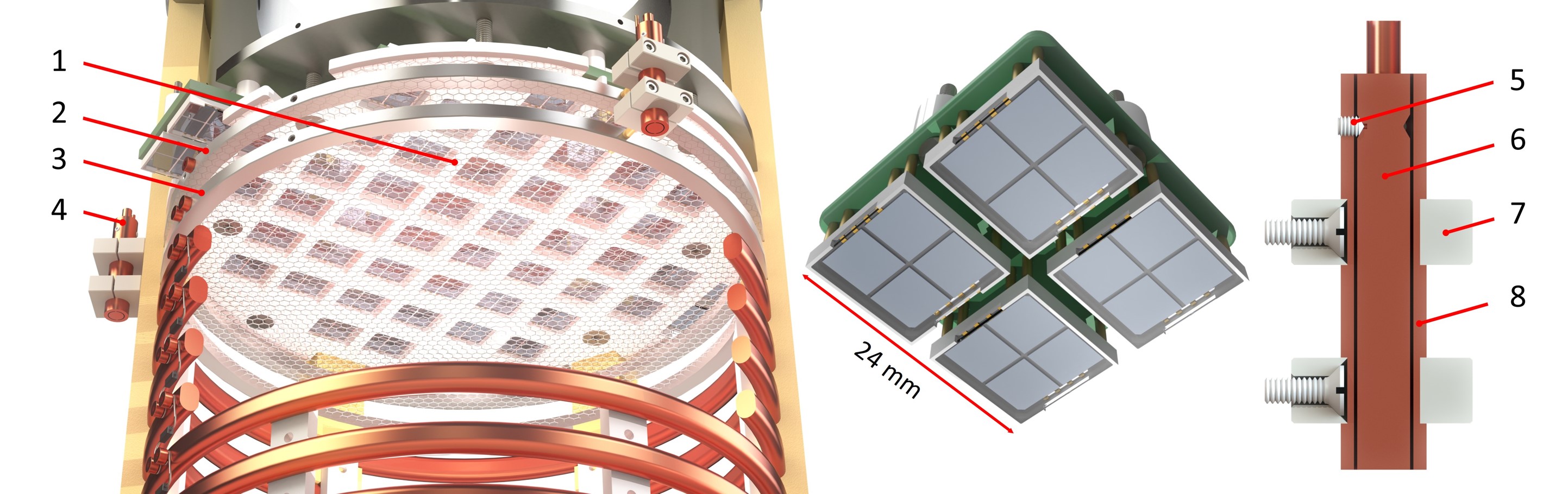}
\caption{(Left): Upper section of the TPC, showing the MPPC array. Legend: (1)~MPPC array; (2)~Anode mesh; (3)~Gate mesh; (4)~Short level meter. (Middle): MPPC tile containing four quad modules of SiPMs. (Right): Short level meter. Legend: (5)~Positioning screws; (6)~Inner cylinder; (7)~Mounting holders, (8)~Outer cylinder.}
\label{fig:TPC_design}
\end{figure}

The liquid level is monitored by three capacitors with a linear dependence of their total capacitance with respect to the LXe level. The working principle of level meters is defined by the difference in the relative permittivity for LXe and GXe (1.96 and 1, approximately)~\cite{Amey:1964}. They consist of concentric cylinders of \SI{4}{mm} inner and \SI{4.5}{mm} outer diameter and \SI{30}{mm} length, similar to the design discussed in~\cite{Hu:2013qdn} and can measure the liquid level and the inclination of the TPC with a dynamic range of a few~$\si{mm}$. Three PTFE set screws centre and level the inner cylinder inside the outer one. The capillary rise of the LXe between the cylinder walls introduces an offset in the liquid level sensor. This factor is considered when finding an optimal ratio between the two cylinder radii, since both the capacitance and the capillarity increase with decreasing distance between the walls. The readout of the capacitors is performed by means of a Universal Transducer Interface (UTI) chip. The capacitors connect to the signal feedthrough via RS-232 cables, and from there to the UTI evaluation board, which has a built-in micro-controller to perform the signal evaluation of the capacitances. This allows for a measurement when there is an unknown gain and offset and provides an auto-calibrating system. A fourth level meter is placed in the gas phase to measure the reference capacitance. 

\begin{figure}[ht!]
\centering\includegraphics[width=0.6\textwidth]{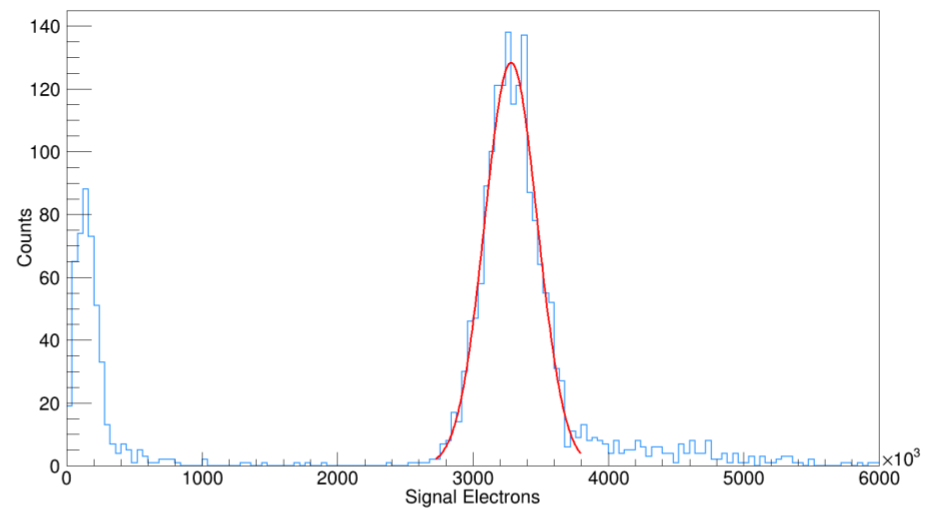}
\caption{SPE gain calibration for a tile with no gain matching between quad modules. The red curve shows a Gaussian fit to the SPE peak. Data were collected at \SI{190}{K} in gaseous argon using $\SI{e4}{events}$ from an external trigger pulsed at \SI{1}{kHz} and dark count is recorded. A gain of $3.3\times10^6$ and a SPE resolution of \SI{6}{\%} (sigma/mean) are estimated from the fit.}
\label{fig:SiPMTileTest}
\end{figure}

The proportional light signal will be observed with an array of 48 \textit{Hamamatsu} S13371 \mbox{$12\times\SI{12}{mm^2}$} VUV4 MPPC installed in the gas phase. We characterised this type of MPPC in~\cite{Baudis:2018pdv} and later used it as a top array in a small LXe TPC, Xurich, as described in~\cite{Baudis:2020nwe}. Each \mbox{$12\times\SI{12}{mm^2}$} "quad" detector is composed of four independent \mbox{$6\times\SI{6}{mm^2}$} MPPCs. We group sixteen MPPC channels (4 quads) into a single channel to form a ``tile''. The array contains 12 tiles, with an effective area of \mbox{$\sim 24 \times\SI{24}{mm^2}$} per tile, and covers $\SI{34}{\%}$ of the area described by the anode, as shown in figure~\ref{fig:TPC_design}. The architecture of the readout board is based on the design from~\cite{Arneodo:2017ozr}, in which a cryogenic readout for sixteen \mbox{$3\times\SI{3}{mm^2}$} VUV4 MPPCs is described. Tiles are mounted on a steel backing disk which is affixed to the pillars of the TPC. Quad modules are connected to the PCB base with push-pin connectors. The array is covered with a PTFE cover plate to mechanically secure the MPPCs. A gain of $3.3\times10^6$ and a resolution of \SI{6}{\%} was measured for a single tile \SI{6}{V} above breakdown voltage and in \SI{190}{K} gaseous argon, as shown in figure~\ref{fig:SiPMTileTest}. This single photoelectron (SPE) resolution is an improvement over the PMTs used in XENON1T~\cite{Antochi:2021buz}. MPPCs will be gain matched per tile to further improve the SPE resolution. The high dark count rate (\SI{0.7}{MHz}) of the MPPC array is not an issue for Xenoscope because the array will be triggered externally, thus avoiding accidental triggers from dark counts. Additionally, the array will observe large signals from the proportional light generated in the gas phase (\SIrange{e2}{e4}{PE}), rendering the small contribution from dark count ($\sim \SI{10}{PE}$) compared to the total detected light negligible. The strong illumination from the ionisation signal will translate to a high dynamic range requirement, past the capability of the v1730SB ADC. To counter this effect, the photosensor array can be operated at a lower gain. The signals can as well be fanned-out to an analog attenuator and can be read in a separate ADC in parallel. The operating over-voltage and corresponding gain will be optimised during commissioning of the array, at the start of the first dual-phase TPC run.

\subsection{High voltage feedthrough}
\label{sec:HV}

The electrostatic field in Xenoscope is produced by biasing the cathode of the $\SI{2.6}{m}$ TPC at $\SI{-50}{kV}$ ($\sim \SI{190}{V/cm}$ drift field). In Xenoscope the baseline design is an \mbox{off-the-shelf} ceramic feedthrough which enters the TPC in the LXe phase via the bottom flange of the inner vessel. Compared to a top-down feedthrough, such as used in the XENON1T and XENONnT experiments~\cite{Aprile:2017aty,XENON:2020kmp}, this results in a shorter connection path to the cathode inside the inner volume, while limiting the distortion of the drift field to a minimum. 

\begin{figure}[t!]
\centering
\includegraphics[width=.3\textwidth]{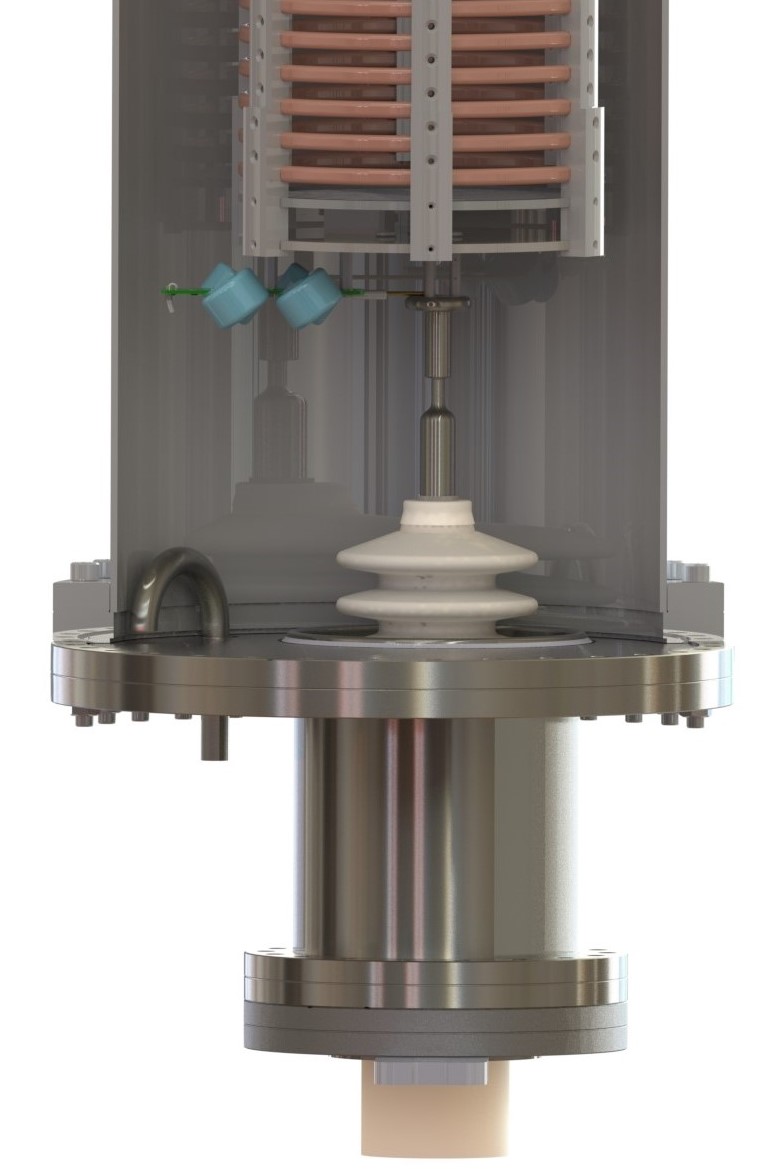}
\includegraphics[width=.35\textwidth]{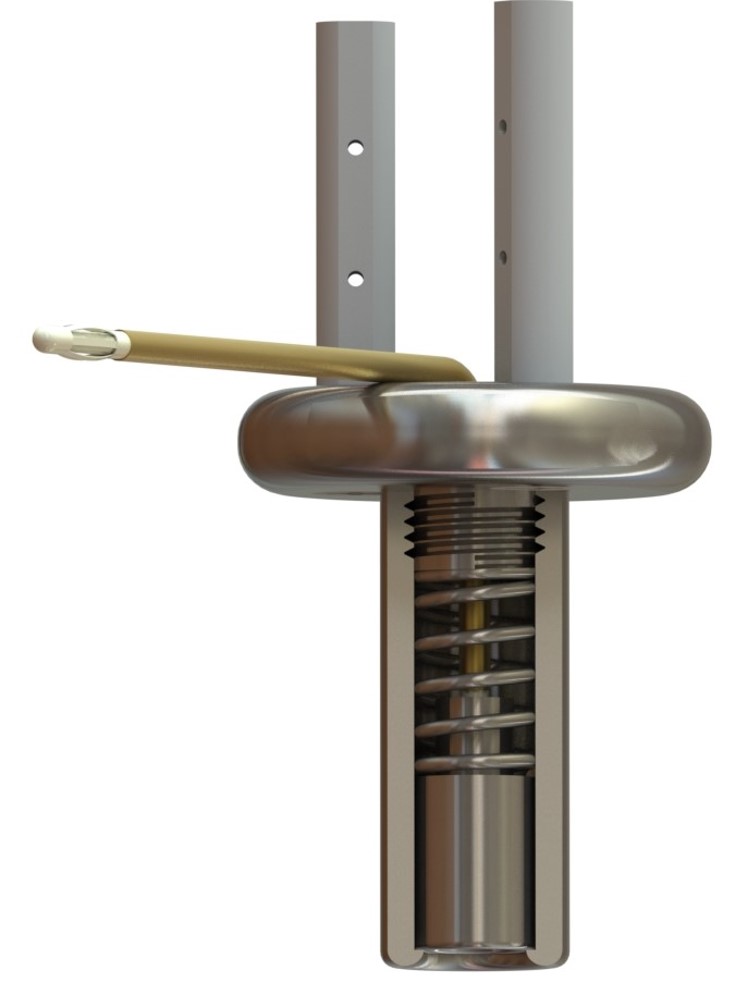}
\caption{(Left): Ceramic HV feedthrough installed on the bottom flange of the inner vessel. (Right): Cup and spring assembly for ensuring HV conductor positioning.}
\label{fig:HV-Assembly}
\end{figure}

The selected \textit{CeramTec} 304 stainless steel and alumina ceramic feedthrough~\cite{CeramTec} is rated up to $\SI{100}{kV}$ at a temperature of $\SI{4}{K}$ and $\SI{8.6}{bar}$ over pressure. This feedthrough is welded onto a DN150CF nipple which extends from the bottom of the inner vessel into the insulation vacuum (figure~\ref{fig:HV-Assembly}, left). A thin rod is threaded onto the feedthrough in order to extend the HV connection into the LXe volume. This rod compresses a cup and spring mechanism which is mounted to the underside of the cathode using PTFE spacers and screws (figure~\ref{fig:HV-Assembly}, right). The cup is designed to stabilise the rod in the case of mechanical misalignment in $(x,y)$, while the spring compensates for height discrepancies in the coupling. Additionally, the spring allows for a constant pressure to be applied on the rod/cup union in order to ensure good electrical contact. An unshielded polyimide cable with a \SI{2}{mm} plug is inserted onto the backside of the cup and is then connected to the charge readout electronics (figure~\ref{fig:circuit}). To bring HV from the air side to the bottom of the cryostat, a second un-shielded polyimide HV cable is run in vacuum from a short cryofitted feedthrough mounted at air-side top flange (on which the Heinzinger power supply plug is connected) to the liquid feedthrough at the bottom. The cable is routed along the length of the inner vessel, guided by insulation brackets.

\begin{figure}[ht!]
\centering
\includegraphics[width=1\textwidth]{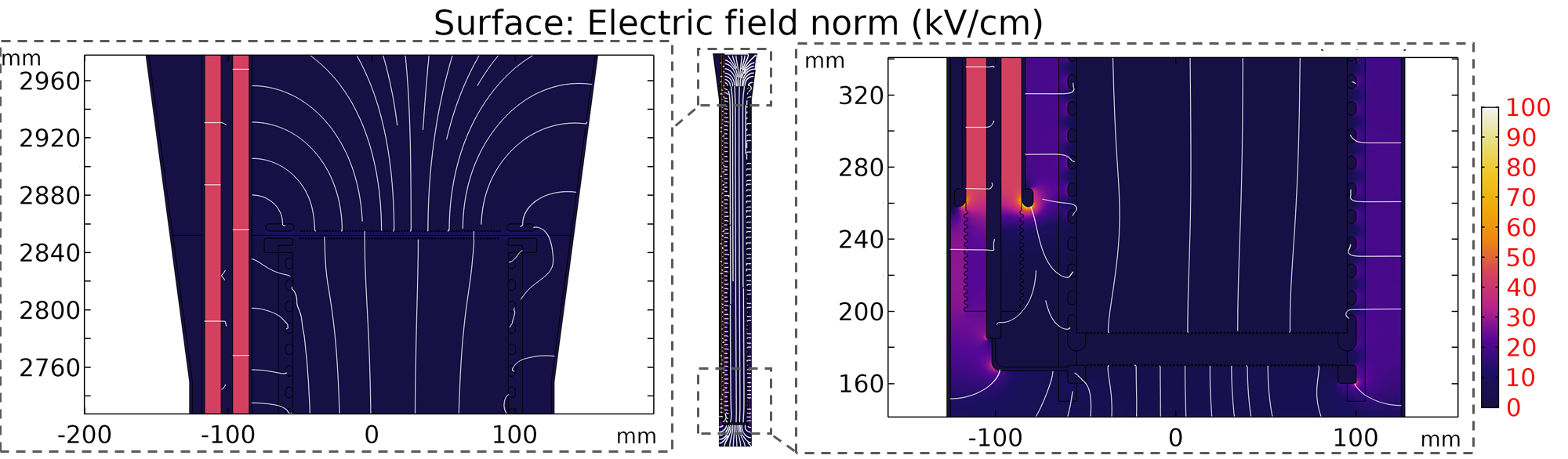}
\caption{Implementation of the TPC design inside the cryostat in COMSOL Multiphysics. The magnitude of the electric field inside the drift region is optimised to be homogeneous along the $z$-coordinate inside the TPC. The colour gradient indicates the norm of the electric field inside the TPC. All regions are below the chosen $\SI{80}{kV/cm}$ limit for the electric field norm. (Left): Zoom onto the top section of the TPC, where the extraction field, GXe/LXe, electrodes (gate and anode), rings and top conical section of the cryostat were modelled. (Middle): Overview of the full \SI{2.6}{m} TPC. (Right): Zoom onto the bottom region in the COMSOL Multiphysics model, where the termination of the HV and the shape of electrodes carrying HV are optimised in order to reduce the magnitude of the electric field, avoiding sharp edges.}
\label{fig:comsol_model_1}
\end{figure}

Although suitable for Xenoscope, ceramic feedthroughs have been shown to be too radioactive for ultra-low background applications, as required in dark matter detectors~\cite{Aprile:2011ru}. Therefore, our facility will also be used to develop fabrication techniques and test \mbox{low-radioactivity} HV feedthrough prototypes suitable for DARWIN. A promising design is similar to the one used in XENON1T/nT, described in~\cite{Aprile:2017aty}, which consists of a stainless steel inner conductor, an ultra-high molecular weight polyethylene (UHMWPE) insulator and stainless steel ground conductor, cryofitted together. A cryofitted feedthrough is not suitable as a liquid feedthrough, for the temperatures of LXe could undo the cryofitting process. Thus, this type of design requires HV entry from the top of the TPC, extending along its length to connect to the cathode at the bottom. To this aim, we constructed a cryofitting station in which we can produce HV power rods of up to \SI{3.4}{m} in length. In anticipation for the testing of HV rods, the PM and TPC will be installed off-centre in the cryostat, ensuring sufficient spacing between the HV rod, the cryostat walls and the detector to avoid electrical discharge, as well as a uniform potential drop among field shaping rings and their shape optimisation. To prevent dielectric breakdown, we decided on an upper limit on the electric field norm of $\SI{80}{kV/cm}$ and optimised the shape of the ground termination around the HV rod based on previous studies~\cite{Cantini:2016tfx,Rebel:2014uia,Tvrznikova:2019xcg}.
Preliminary electric field simulations were performed with the COMSOL Multiphysics\textsuperscript{\textregistered} simulation software~v.5.4~\cite{multiphysicscomsol}, shown in figure~\ref{fig:comsol_model_1}, to ensure homogeneity of the electric field and to optimise the positions of the detector and the HV rod entering from the top flange through the GXe phase. We expect the hydrostatic pressure from the LXe column to suppress bubble formation in the LXe, as seen in~\cite{Bay:2014jwa}. As an extra measure, we set a trip current on the HV supply of $\SI{1}{\mu A}$ to prevent electrical breakdowns caused by potential GXe bubble formation. Future tests with low pass filters to cut potential ripple noise from the HV power supply are being considered.

\subsection{Data acquisition system for the TPC}
\label{sec:DAQ}

Waveforms from the PM and the MPPC array will be digitised using a CAEN~\cite{CAEN} V1730SB, 16 channel, $\SI{500}{MS/s}$ analog to digital converter (ADC) with $\SI{5.12}{MS/channel}$. The ADC is connected to a Linux machine through fibre optics and a PCIe optical link (CAEN A3818B). It is programmed using custom C++ software that relies on CAEN libraries which have functions for high-level management. Data is read out from the ADC card by performing a block transfer of data using proprietary software from CAEN. This software also allows for live time plotting of the integrals of baseline-subtracted waveforms. In the case of gain calibrations for the MPPC array, the SPE peak can be fitted, and the mean and width can be monitored. 

The S2 signal is expected to be detected $\sim \SI{1.7}{ms}$ after the trigger provided by the flash lamp. In order to save disk space, an internal software delay is used. The buffer on the ADC allows for waveforms of up to \SI{5}{ms} to be stored per readout cycle. The acquisition window size for both trigger modes will be tuned to match the achieved electric field strength and electron drift speed. 

The data is stored as a compressed ROOT file \cite{root} which contains the raw ADC counts per digitisation. The ADC configuration file is also saved for every acquisition. The current data analysis stream and framework is based on the processes described in~\cite{Baudis:2020nwe}. The pulse-finding algorithm has been adapted from~\cite{Zugec:2016gen}, which uses a smoothed derivative method to identify signal pulses as large ($3.5 \sigma$) fluctuations in the noise. Four threshold crossings of the derivative define a pulse. Local minima are identified using the derivative allowing individual pulses to be distinguished in the event of pulse pileup.

\section{Slow control system}
\label{sec:SC}

The slow control (SC) system  for Xenoscope was developed to provide means for monitoring and control of the experiment. It also provides the user with notifications in case of anomalies or emergencies. 
The SC system is designed with a micro-service architecture~\cite{microservice} and combines well-established automation software with state-of-the-art orchestration, monitoring and alarming tools usually used in the context of large-scale web services.
It consists of several independent subsystems:

\begin{figure}[ht!]
\centering
\includegraphics[width=1\textwidth]{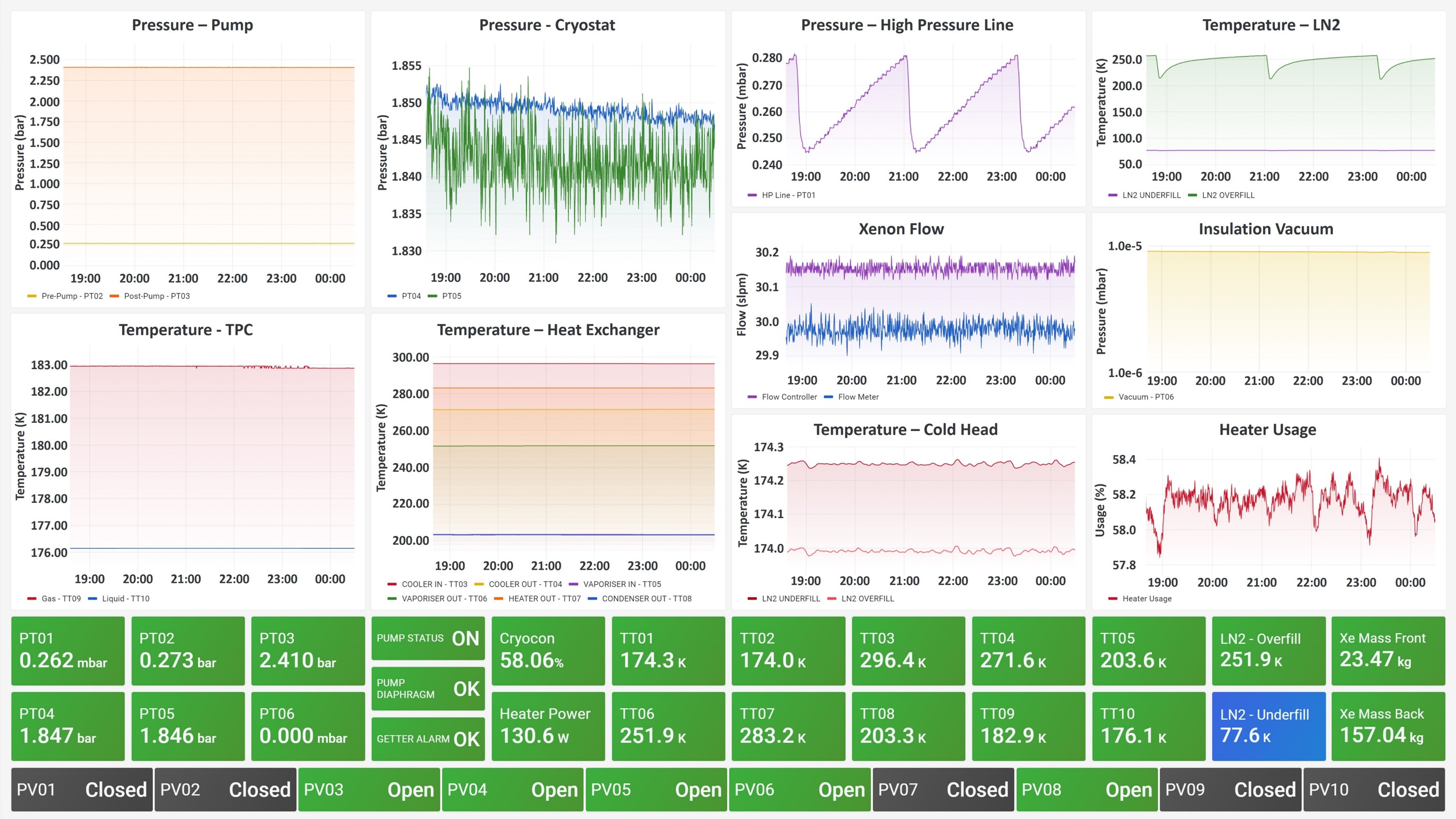}
\caption{Dashboard of the slow control system in Grafana. The customised graphical user interface displays data over a user-defined time range and in real-time. Dashboards are also used to set alarm thresholds.}
\label{fig:SC-dashboard}
\end{figure}

\begin{figure}[ht!]
\centering
\includegraphics[width=0.8\textwidth]{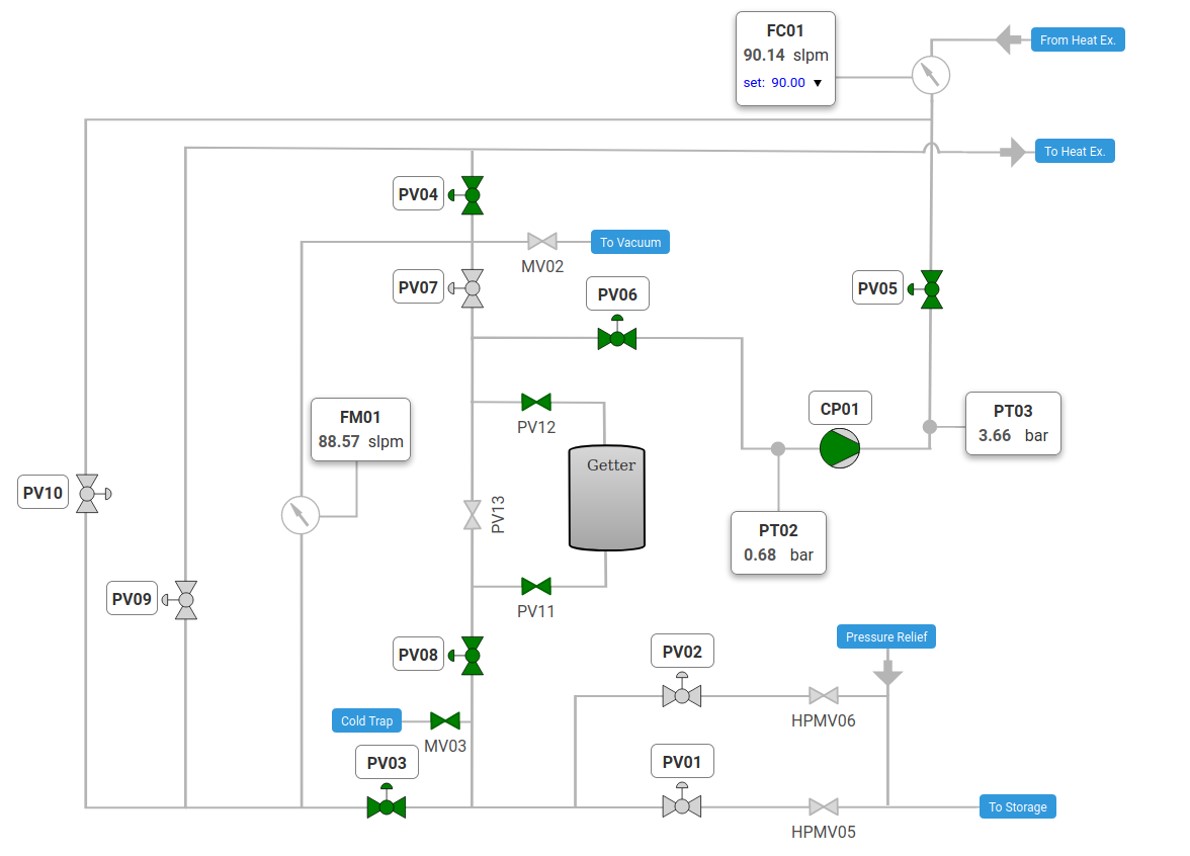}
\caption{Graphical user interface of the system. 
The control panel of the gas system of the experiment is shown as an example of a control screen.}
\label{fig:SC}
\end{figure}

\begin{itemize}

    \item {\textbf{Micro-controllers:} These devices are responsible for collecting sensor data and for controlling hard-wired actuators (e.g.~valves, pumps, flow controllers).
    In terms of hardware, the micro-controllers consists of an industrial Programmable Logic Controller (PLC) and one Raspberry~Pi~\cite{rasp}. The data they collect are then exposed to the rest of the system through open network protocols, namely Open Platform Communications (OPC)~\cite{opc} and HTTP.}
    
    \item{ \textbf{Data ingestion:} Data collected from each sensor is ingested into Prometheus~\cite{prom}, a well-established time-series database dedicated to monitoring and alarming.}
    
    \item{\textbf{Visualisation platform:} 
    Grafana~\cite{graf} is used for the monitoring of the acquired data. This tool also allows for the setting of alarm thresholds for any monitored time series. 
    An example of an end-user graphical interface is shown in figure~\ref{fig:SC-dashboard}.
    }
    
    \item{ \textbf{Control platform:}
    We designed two open-source tools that allow for bidirectional data flow between micro-controllers and the user interface. The first one is a proxy service~\cite{opcp} that exposes the OPC server through an HTTP Application Programming Interface (API). The data streamed by the PLC can then be queried with this tool using simple HTTP requests. The second is a front-end framework~\cite{jshmi} used to simplify the design of control screens for browsers. An example of these control screens is shown in figure~\ref{fig:SC}.}
    
    \item{\textbf{Alarm and notification system:}
    An alarm system, part of a set of tools offered by Prometheus, continuously processes the collected data searching for anomalies based on configured rules. Grafana can be used to visually set these alarm thresholds and rules with a Graphical User Interface, which are then used to trigger alarm notifications with Prometheus. 
    A third-party notification system takes care of notifying users via e-mail and SMS.
    }
    
    \item{\textbf{Micro-service orchestration:}
    Each service described above, except for the micro-controller software, is broken down into isolated Linux containers~\cite{docker}, following best practices of micro-service architecture. These micro-services are then orchestrated by Kubernetes~\cite{kube}, an open-source container-orchestration system for automating computer application deployment, scaling, and management. Kubernetes ensures that all services are always running. }

\end{itemize}
 All used software is open-source, apart from the proprietary PLC programming tools. The system is designed to be fault tolerant and, as much as possible, secure.

\begin{figure}[h!]
\centering
\includegraphics[width=0.8\textwidth]{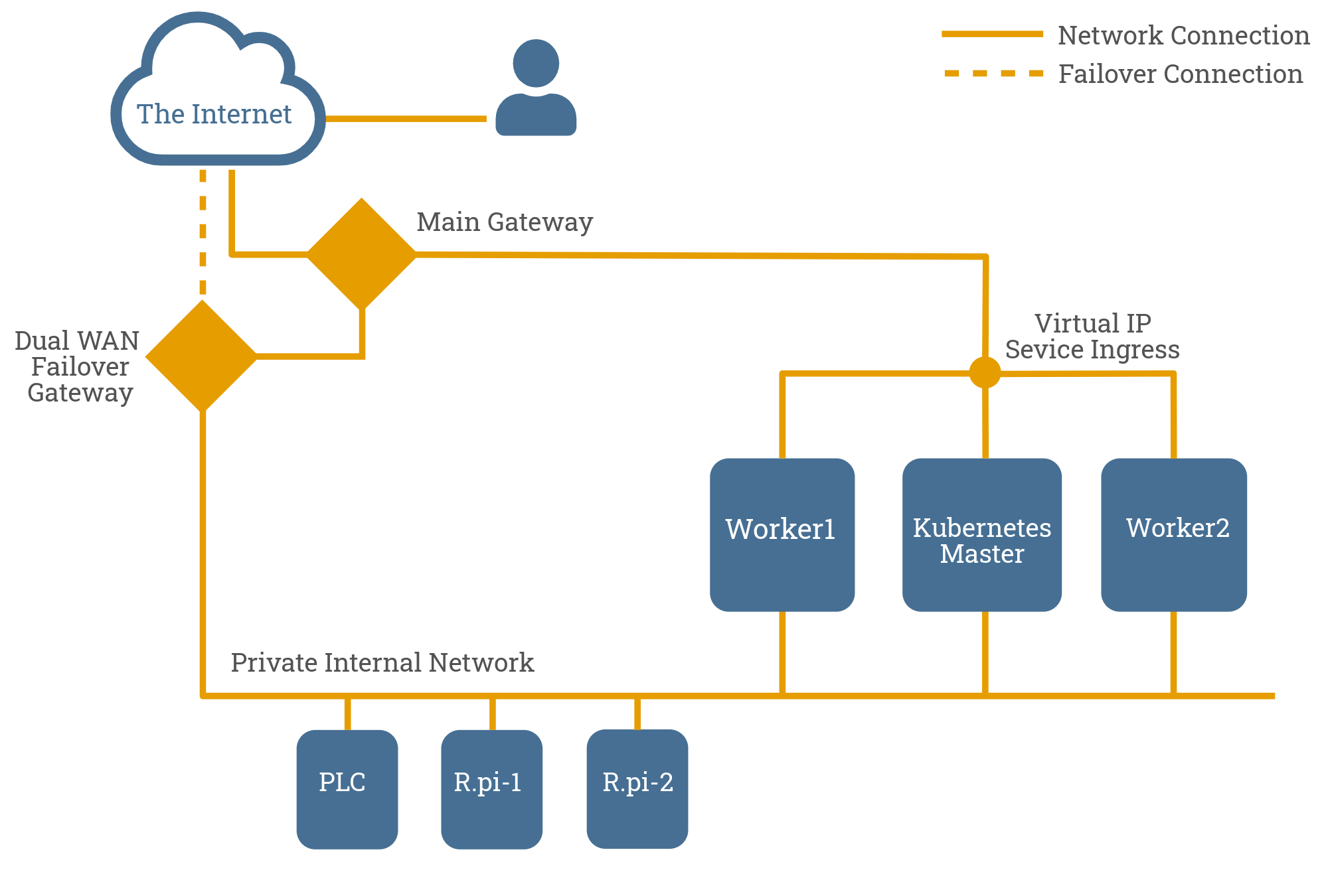}
\caption{Networking configuration of the monitoring and control system. Three servers are operated in a Kubernetes master-worker configuration; they are connected to the internet via the main University gateway and a secondary dual Wide Area Network (WAN) fail-over router. The micro-controllers are instead connected only to an internal private network and not accessible from the internet.}
\label{fig:SCLayout}
\end{figure}

One of the main challenges of a monitor and control system is its reliability, especially if it must guarantee the safety of the operators and of valuables such as xenon. 
By ``reliable system'', we intend a system that can tolerate and automatically react to single points of failure, without manual intervention from the user and without loss of service.
A common measure of the reliability of a service is its availability over a year in percent. We aim to reach $\SI{99.99}{\%}$, which corresponds to about $\SI{50}{min}$ downtime per year, including maintenance.
The key to achieve such high availability is redundancy. The system is thus designed with redundant hardware and network connections, while the micro-service architecture guarantees software failure isolation, meaning that a failure of one service does not affect the others.

As described above, we use Kubernetes as orchestration software for micro-services. It ensures that the system is healthy by restarting services in case they become unresponsive. We follow a master-worker type of architecture with three machines: one master and two workers. The master node continuously checks the availability of each service and each worker node. In case one of the workers becomes unavailable, the load is automatically transferred to the other one. In case the master becomes unavailable, there is no effect on the provided service. The networking configuration of the system is shown in figure~\ref{fig:SCLayout}.

The dual-WAN fail-over router provides network redundancy. The default network service is provided by the University. In case of failure, the router automatically switches to a Long-Term Evolution (LTE) network service. In this last condition, the system is not directly reachable. However, it will send alarm notifications. Furthermore, an external server receives continuously (every few minutes) ``heartbeats'' from the system, alerting the users in the unlikely event of it being incapable of sending alert notifications.

\section{Facility commissioning}
\label{sec:commissioning}

\begin{figure}[ht!]
\centering
\begin{minipage}{0.48\textwidth}
\centering
\includegraphics[width=1\textwidth]{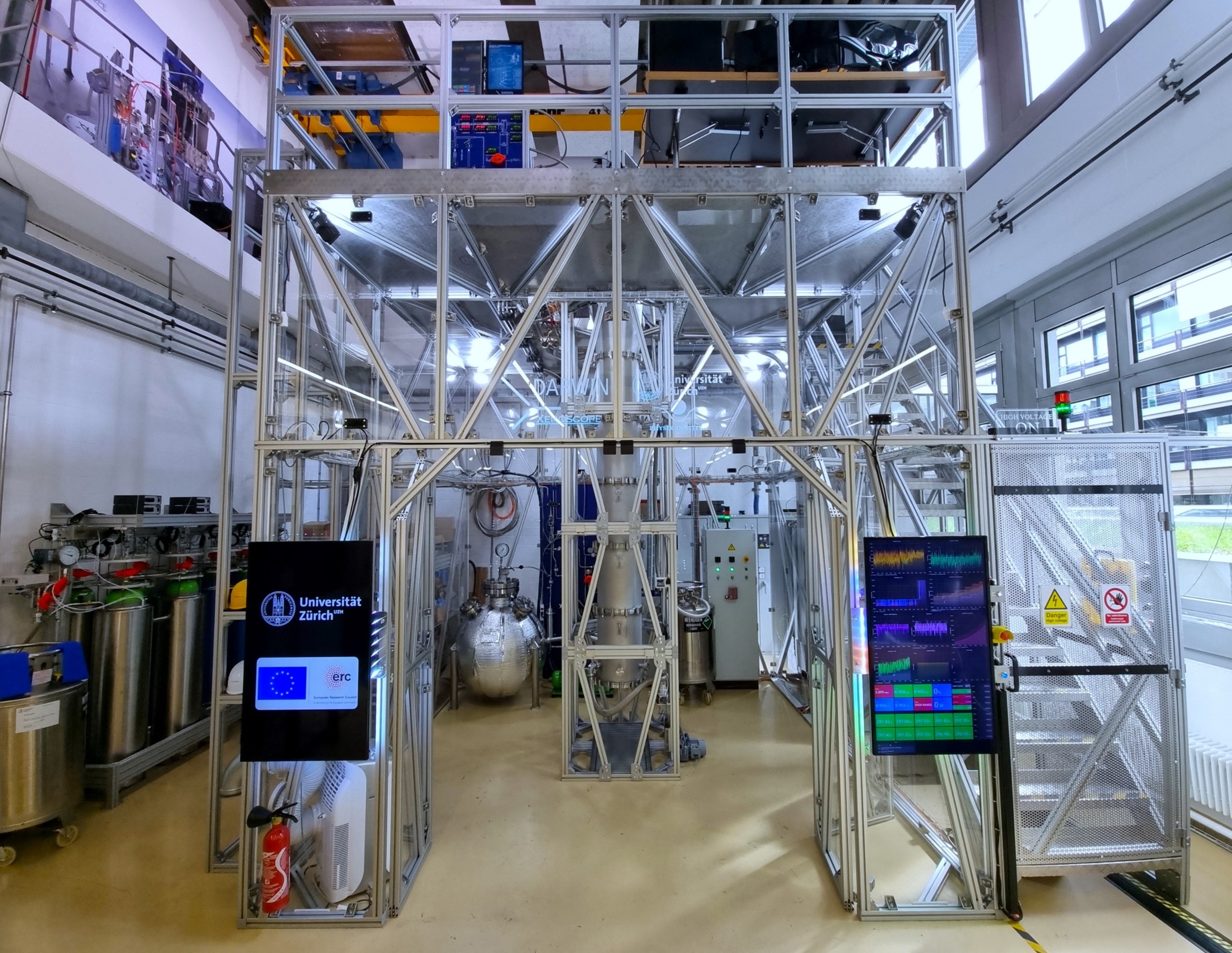}
\end{minipage}\hfill
\quad
\begin{minipage}{0.48\textwidth}
\centering
\includegraphics[width=1\textwidth]{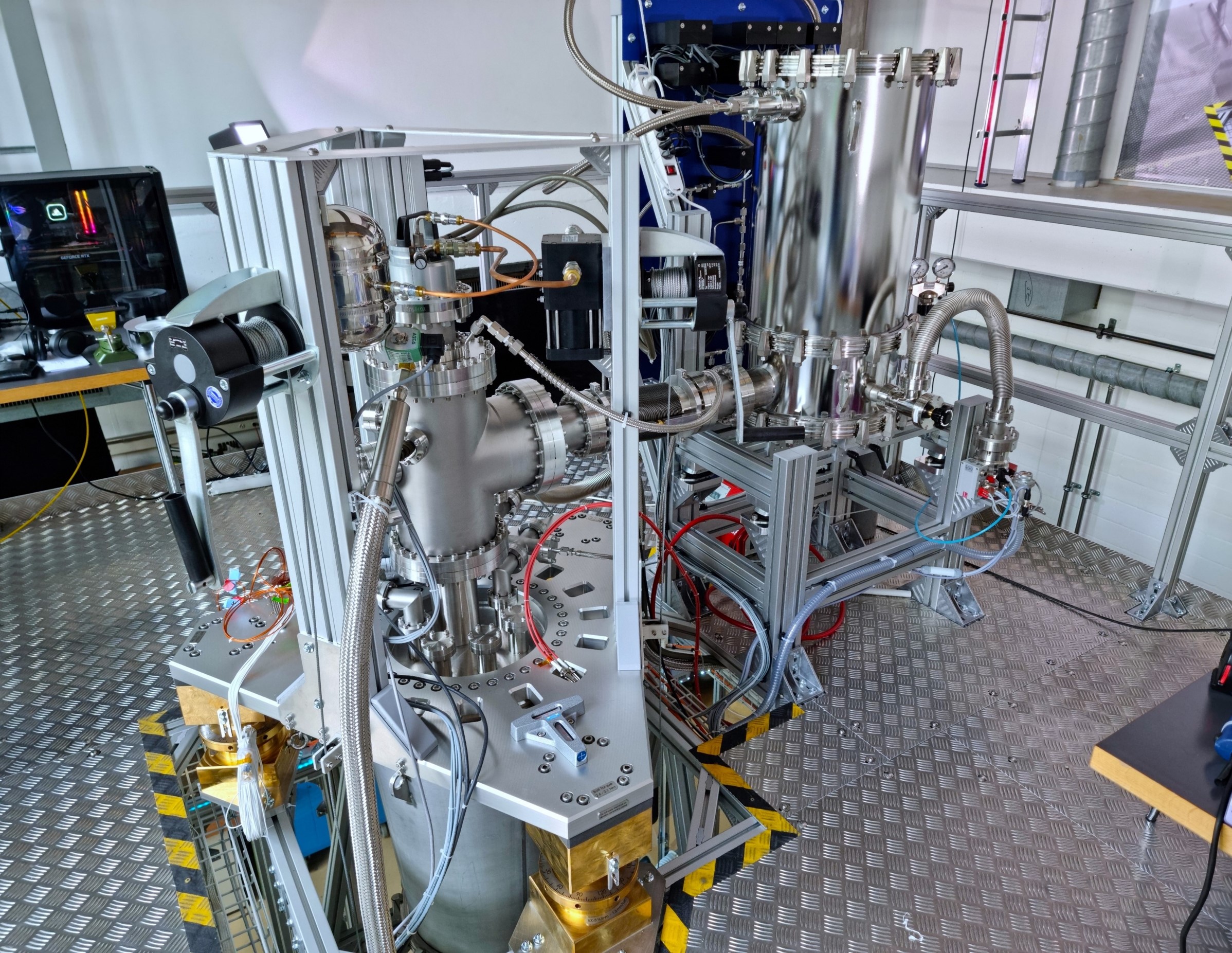}
\end{minipage}
\caption{(Left): Picture  of the facility as seen from the front. Visible are the support structure, the outer cryostat and the xenon storage and recuperation systems. (Right): Picture of the top of the facility, showing the cryogenic system.}
\label{fig:pictures}
\end{figure}

In this section we describe the commissioning phase of the Xenoscope facility which began in early 2021 and the first run with LXe following the completion of the construction phase, as seen in figure~\ref{fig:pictures}. The goal of this test phase was the verification of the functionality, reliability and safety of all subsystems and the benchmarking of their performance. In particular, the maximum achievable recirculation flow, which is limited by the pressure rating of the xenon compressor, and the heat exchange efficiency have been evaluated.

The commissioning started with leak and pressure testing of the gas system. The entire gas system and cryostat assembly (only 1/3 of the full cryostat height) were first leak-checked with maximum leak rate of \SI{e-9}{mbar\cdot l/s}. The high-pressure side of the gas system was pressure tested at \SI{55}{bar} and the low-pressure side at \SI{4.2}{bar}.

\subsection{Xenon transfer and recirculation}
The system was evacuated to a pressure of \SI{1e-5}{mbar} and was at first filled with \SI{1.37}{kg} of GXe at \SI{2.1}{bar} to test the filling procedure and the performance of the GXe compressor and getter unit. The GXe was purified for $\sim$ \SI{44}{h} at room temperature and was then recuperated into the bottle storage array as proof of concept. During filling, the xenon was recirculated at \SI{22}{slpm} in the purification loop to increase the convection in the cryostat which is necessary for its cool-down due to the lack of natural gas convection between the cooling tower and the cryostat. Xenon from the storage system was introduced in the purification loop through the pressure reducer located after the compressor, allowing for the fine-control of the pressure in the system during the filling procedure. The first LXe fill was performed discontinuously over a period of 5 days with the transfer of $(80.66\pm0.10) \, \si{kg}$ of xenon. After the first run with LXe that lasted for about 5~weeks, the xenon was transferred back to the storage array by cryo-pumping (as described in section~\ref{Sec:Gas_recovery}) continuously over a period of two days at an average rate of $\SI{5}{slpm}$. The system was found to be stable during this process. Within the maximum accuracy of the mass scale system of $\pm \SI{100}{g}$, no loss of xenon could be observed from the commissioning run.

\begin{figure}[ht!]
\centering
\begin{minipage}{0.48\textwidth}
\centering
\includegraphics[width=1\textwidth]{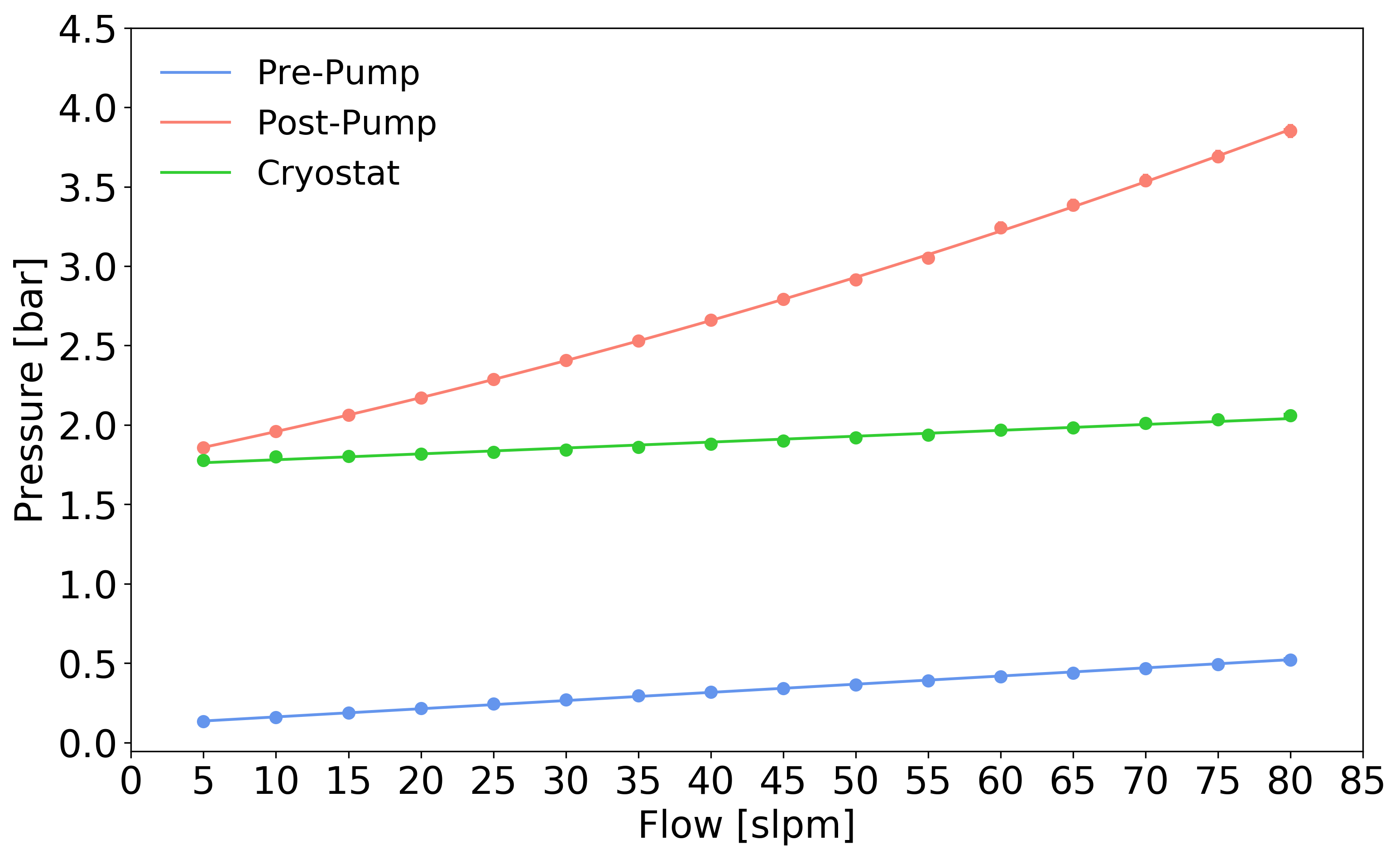} 
\end{minipage}\hfill
\quad
\begin{minipage}{0.48\textwidth}
\centering
\includegraphics[width=1\textwidth]{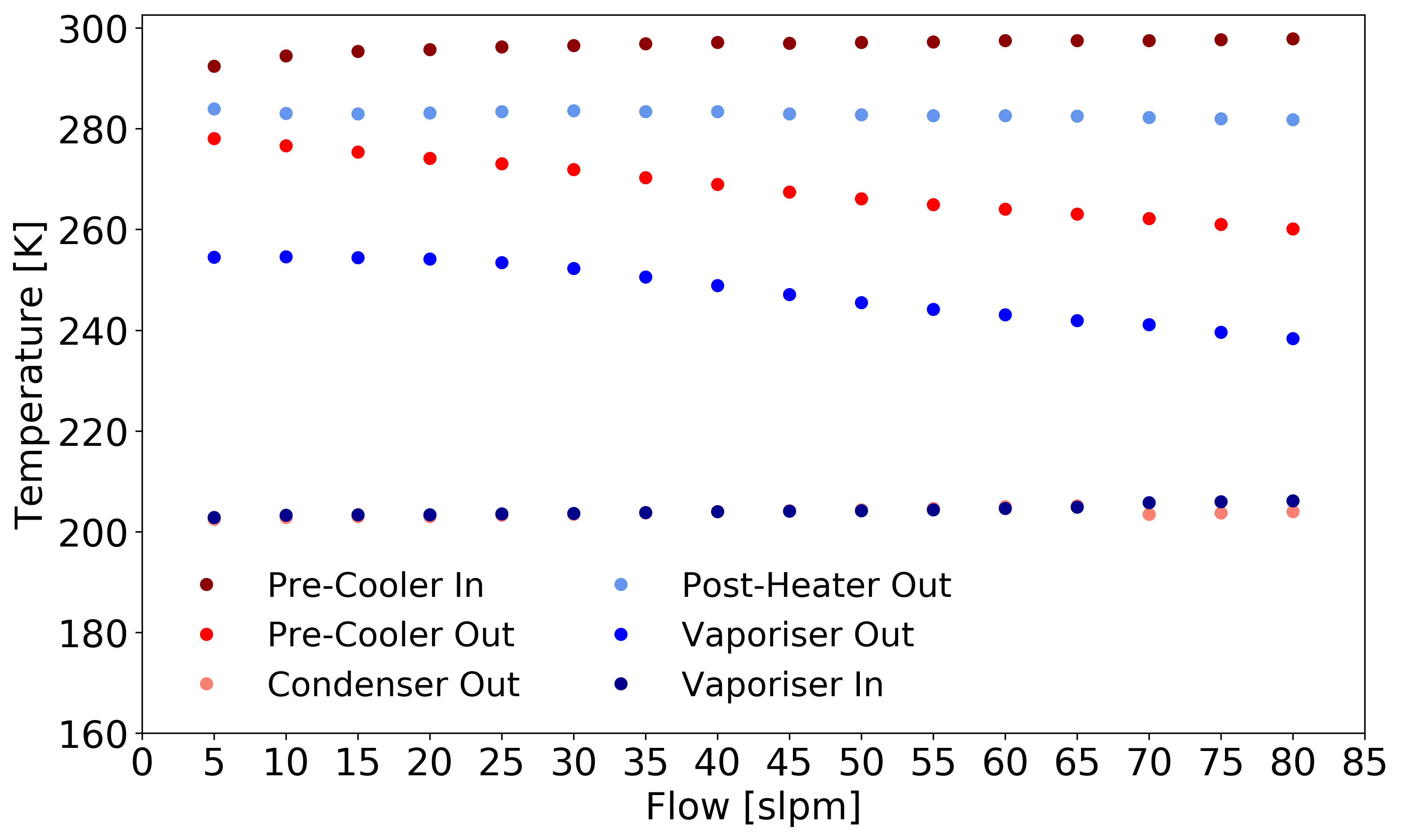}
\end{minipage}
\caption[Flow-dependent xenon compressor pressures and heat exchanger temperatures]{Data at $\SI{174.25}{K}$ cold head temperature set-point (error bars are smaller than the markers). (Left) Pressures at the xenon compressor and inside the cryostat for various flow rates. The fit function used for the post-pump pressure is a second order polynomial. (Right) Temperature hierarchy of the heat exchanger inlets and outlets.}
\label{fig:pump_pressure_he_temperature}
\end{figure}

The purification system was tested at flows ranging from \SI{5}{slpm} to \SI{80}{slpm} with a cold head temperature of \SI{174.25}{K}. Figure~\ref{fig:pump_pressure_he_temperature}, left, shows flow-dependent pressures before and after the compressor, as well as in the cryostat. We deduce from the data that the flow at the maximum rating of the compressor of \SI{4}{bar} absolute is \SI{83}{slpm}.

\subsection{Heat exchange efficiency and thermal leakage}
\label{sec:commissioning_he_efficiency}

Benchmarking during commissioning included the determination of the efficiency of the heat exchange and the evaluation of constant and flow-dependent heat leaks. For all the presented tests, we assume that the supply line close to the opening to the cryostat, and in particular throughout the heat exchangers, has the same constant pressure as the cryostat itself, and that all thermodynamical changes are isobaric. The underpressure in the return line cannot be measured directly with the present instrumentation, and no assumptions on its value are made. In particular, we do not use the enthalpy information of the liquid in the return line. However, pressure uncertainties on the gas side of the heat exchangers do not impact the presented analysis. The maximum (minimum) pressure in the system is present directly after (before) the xenon compressor with a maximal difference to the cryostat pressure at high (low) flows (see figure~\ref{fig:PID}). This difference includes the flow-dependent pressure drop over the getter (flow-controller) on the supply (return) line and the connecting tubing. However, even if the pressure at the pre-cooler inlet (post-heater outlet) would be equal to the post-pump (pre-pump) pressure, the gas enthalpy would only differ by $0.5 \%$ in both cases~\cite{NIST}. Additionally, we assume that the cooling power is constant, and changes of the insulation vacuum as well as of the laboratory temperature are negligible. The cold head set-point was chosen to be $\SI{174.25}{K}$ and kept constant for all flows. This results in a flow-dependent cryostat pressure that directly influences the liquid temperature which is at its boiling point. For this particular temperature, the normal operating pressure of $\sim \SI{2}{bar}$ is obtained at $\SI{70}{slpm}$. All measurements were taken in thermal equilibrium if not mentioned otherwise. To that end, the system was allowed to relax for 24 hours after flow changes such that all pressure and temperature parameters stabilised.

\subsubsection*{Thermal equilibrium and gas enthalpy change}
\begin{figure}[ht!]
\centering
\begin{minipage}{0.48\textwidth}
\centering
\includegraphics[width=1\textwidth]{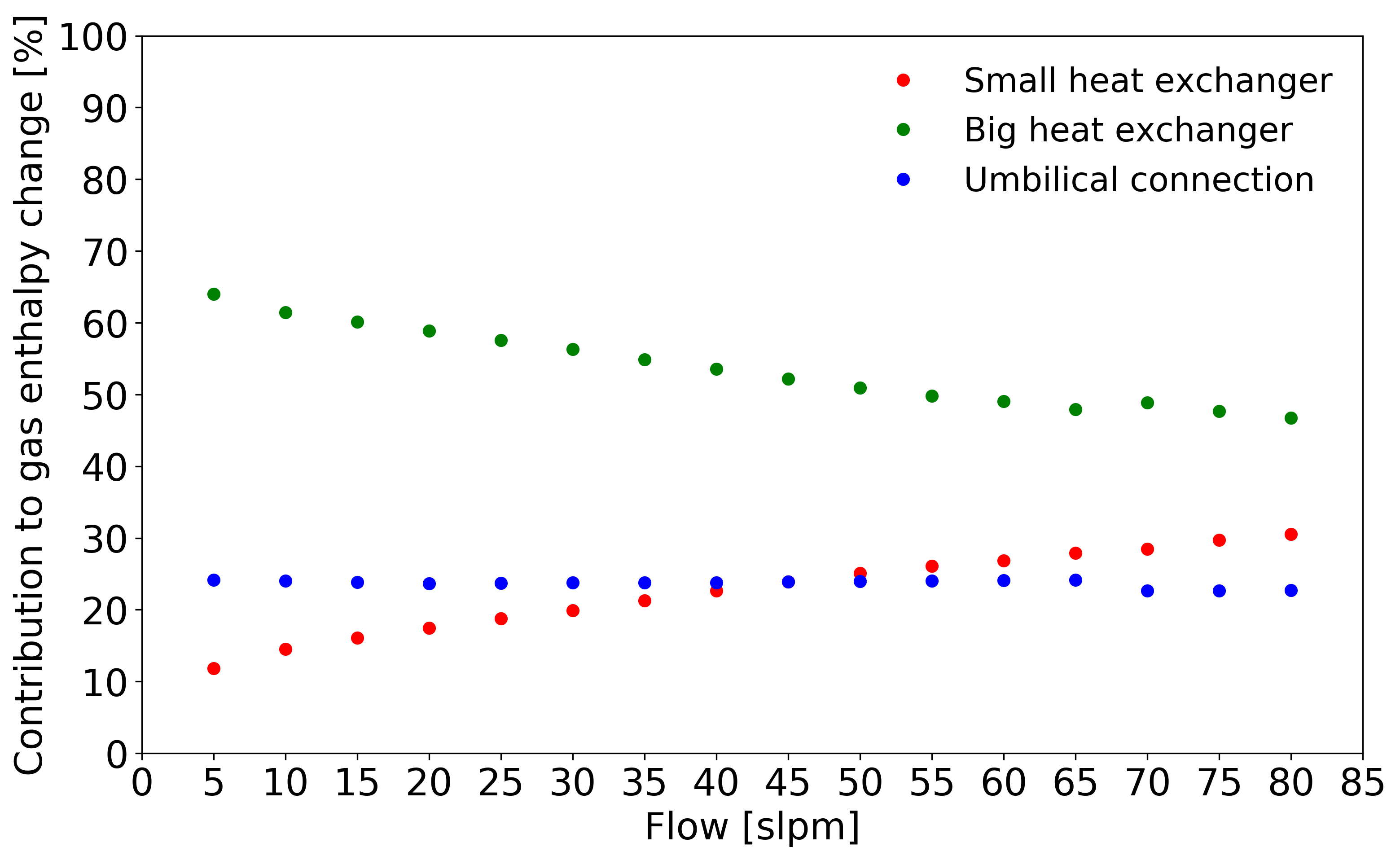} 
\end{minipage}\hfill
\quad
\begin{minipage}{0.48\textwidth}
\centering
\includegraphics[width=1\textwidth]{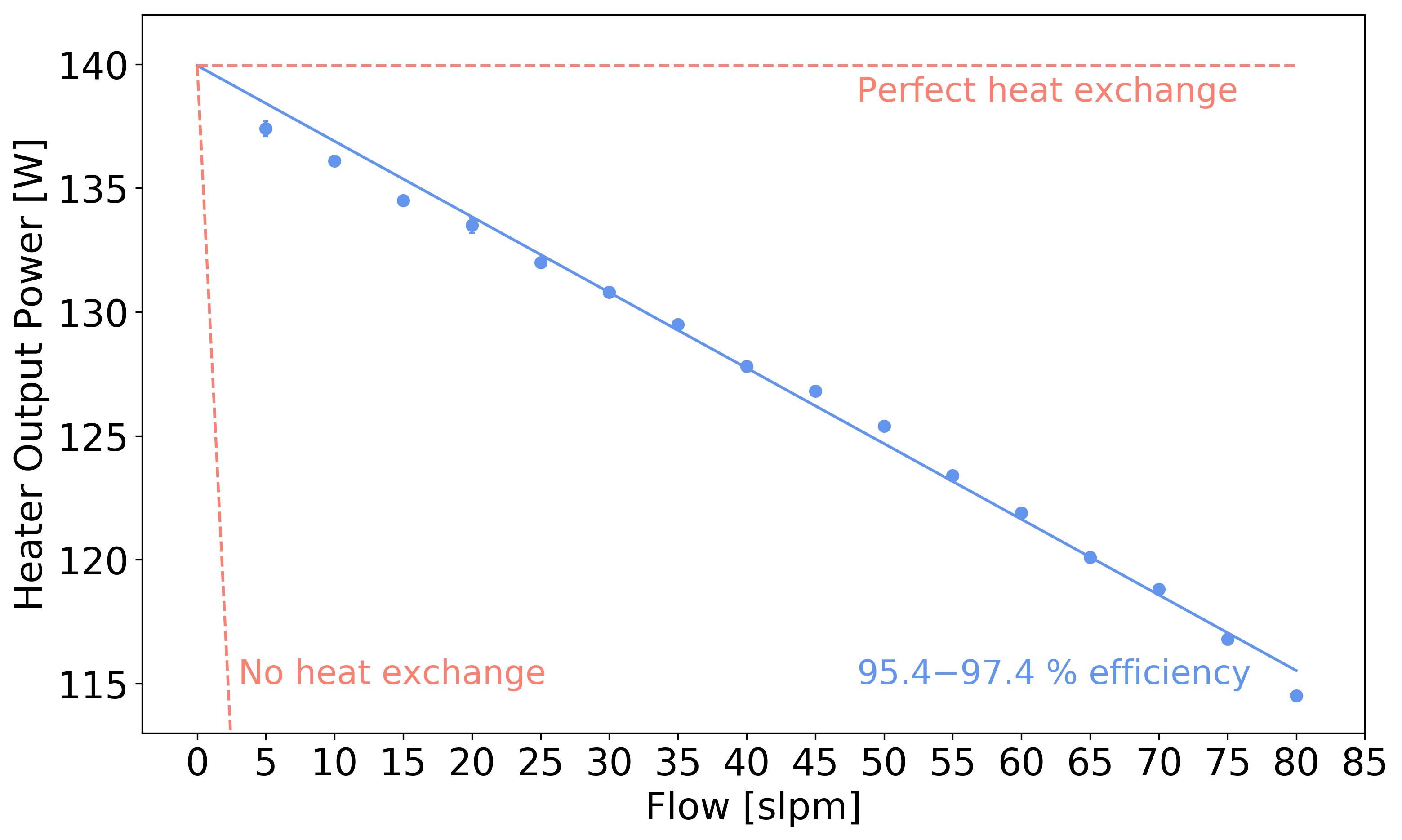}
\end{minipage}
\caption[Contributions to the gas-gas exchange and heater output power]{Data at $\SI{174.25}{K}$ cold head temperature set-point (error bars are smaller than the markers). (Left): Contribution of the heat exchangers to the gas cooling process. (Right): Linear decrease of the heater output power with increasing flow rates.}
\label{fig:gas_enthalpy_contributions_heater_power}
\end{figure}

In figure~\ref{fig:pump_pressure_he_temperature}, right, we show the temperature hierarchy of the inlets and outlets of the heat exchangers that follow the expectation. However, at some flow rates, we observed a lower temperature at the condenser outlet than at the vaporiser inlet. This is attributed to the fact that the temperatures are measured on the connecting tubes close to the nozzles of the heat exchangers. Hence, the expansion of the outgoing gas after the temperature sensor would cool down the bottom of the large heat exchanger and in turn lead to a lower measured temperature on the supply line. We observed that the inlet of the pre-cooler, as well as the condenser out- and vaporiser inlet, became colder at lower flows. The first is due to the smaller heat input from the recirculation pump and the latter is due to the lower liquid temperature. Temperatures higher than the boiling point of xenon at the respective pressures suggest that the gas in the supply line does not condense inside the heat exchangers. The direct observation of LXe at the end of the supply line through the viewport points to a liquefaction in the umbilical connection, which we attribute to the physical contact of the tight MLI packing between the ingoing and outgoing hose. In fact, this increases the heat exchange efficiency. The contributions of the heat exchangers to the gas-gas exchange can be seen in figure~\ref{fig:gas_enthalpy_contributions_heater_power}, left. Depending on the flow, this process only accounts for $\sim 1/5$ of the total isobaric enthalpy change as the majority is inherent to the phase change. 
 
\subsubsection*{Heat exchange efficiency}
The determination of the heat exchange efficiency is based on the idea outlined in~\cite{Aprile:2012jh}. Assuming a constant cooling power at a certain cold head temperature set-point, the power output of the electrical heater on the cold head determines the actual required cooling power for steady state operation. Neglecting any constant and flow-dependent heat leaks that emerge when recirculating, a lower limit on the heat exchange efficiency can be calculated from:
\begin{equation}
\label{eq:he_efficiency}
\epsilon = 1- \frac{P(0)-P(r)}{\Delta H \cdot \rho \cdot r}. 
\end{equation}
Here, $P(0)$ and $P(r)$ $[\si{W}]$ are the power outputs of the heater at zero and some recirculation rate $r$ $[\si{slpm}]$, respectively. The enthalpy change of the xenon from liquid at its boiling point to warm gas is denoted by $\Delta H$ $[\si{W\cdot min/g}]$ and the mass density of xenon under standard conditions by $\rho=\SI{5.8982}{g/l}$. The enthalpy change was calculated individually for all flow rates for the isobaric process of cooling down from the respective pre-cooler inlet temperature to the liquid temperature based on NIST data~\cite{NIST}.

\begin{figure}[ht!]
\centering
\begin{minipage}{0.48\textwidth}
\centering
\includegraphics[width=1\textwidth]{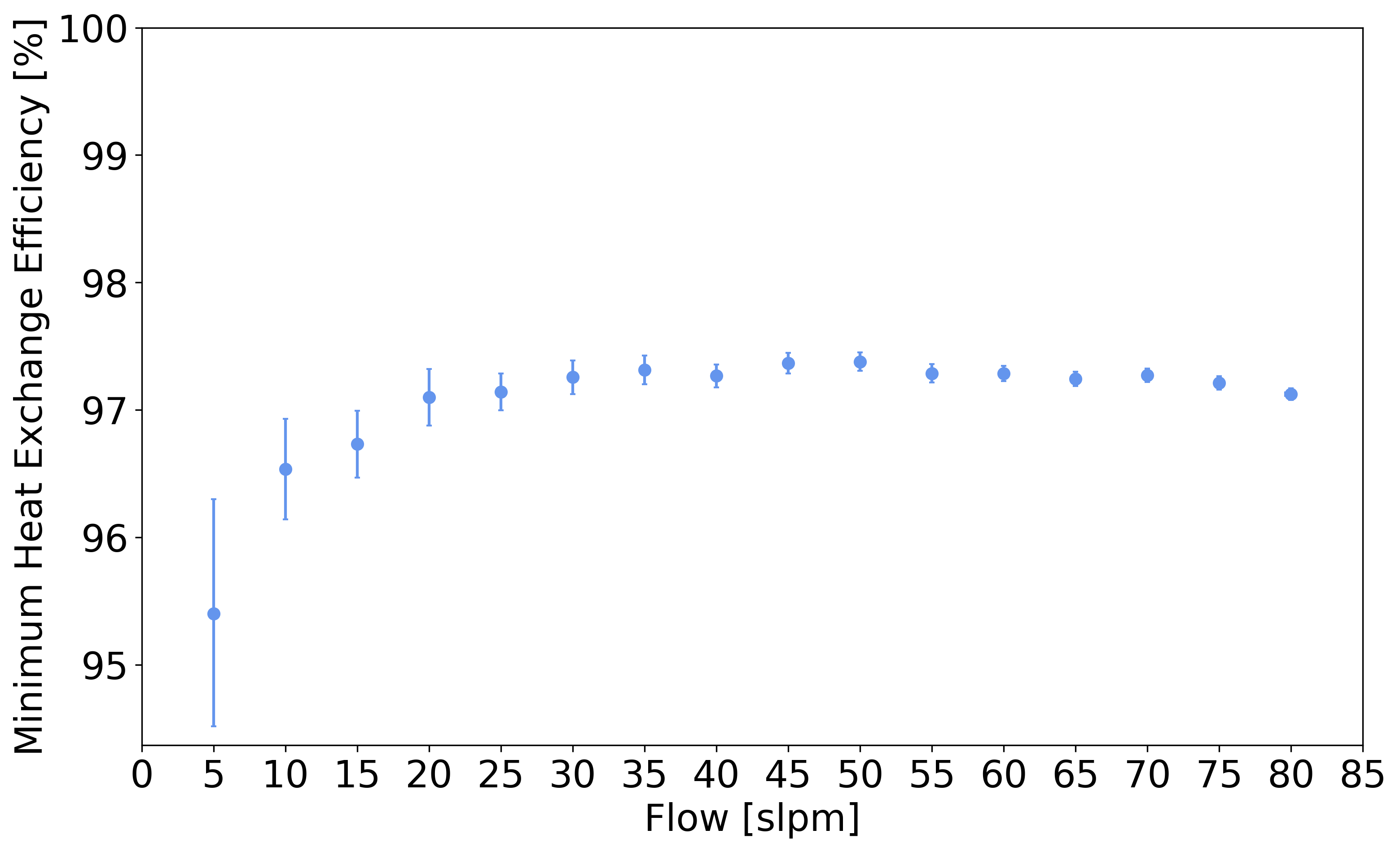} 
\end{minipage}\hfill
\quad
\begin{minipage}{0.48\textwidth}
\centering
\includegraphics[width=1\textwidth]{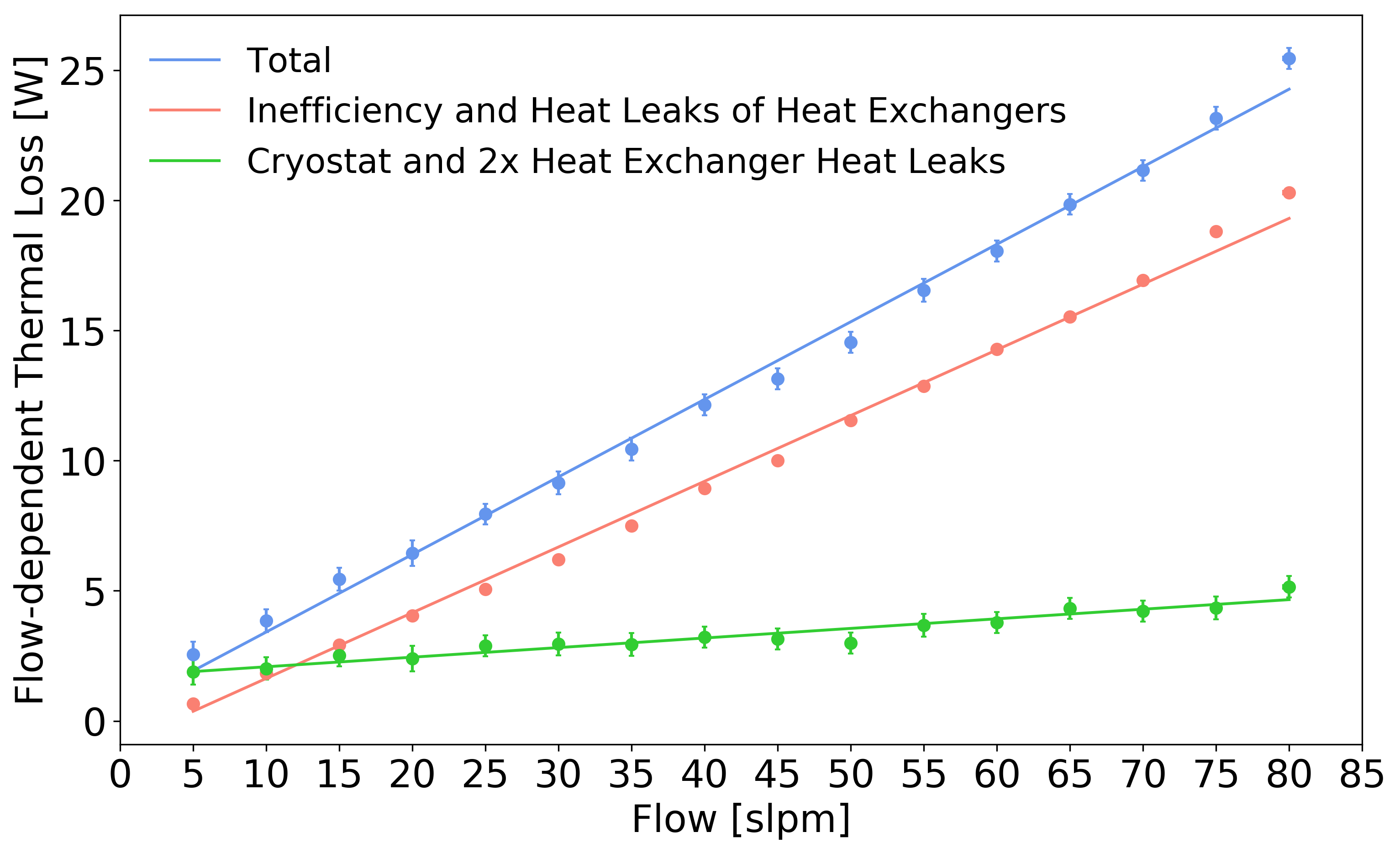}
\end{minipage}
\caption[Heat exchange efficiency and thermal losses]{Data at $\SI{174.25}{K}$ cold head temperature set-point (error bars are partially smaller than the markers). (Left): Lower limit on the total heat exchange efficiency at various flow rates. (Right): Flow-dependent thermal losses fitted with linear regressions.}
\label{fig:he_efficiency_thermal_leakage}
\end{figure}

As expected, we obtain a linear heater output that decreases with flow, see figure~\ref{fig:gas_enthalpy_contributions_heater_power}, right. From the linear regression we obtain a slope of $(0.305 \pm 0.007) \, \si{W/slpm}$ and a power-axis intercept at $(139.9 \pm 0.4) \, \si{W}$ which represents the zero heater output in absence of constant recirculation-based heat leaks. In fact, the actual heater output power at zero recirculation flow is in the range $\SIrange{154}{174}{W}$. This estimate originates from alternating low-flow and zero-flow measurements as, due to the lack of convection inside the cryostat, the system cannot be brought into a steady state below $\SI{2.5}{bar}$ without recirculation.
In figure~\ref{fig:he_efficiency_thermal_leakage}, left, we show the lower limit on the heat exchange efficiency. It is constant at a mean value of $\SI{97.3}{\%}$, in the range $\SIrange{20}{80}{slpm}$. We found a lower efficiency of $>\SI{95.4}{\%}$ in the range $\SIrange{5}{15}{slpm}$. At very low flows of $\sim \SI{1}{slpm}$, the heat exchangers warm up and the efficiency increases rapidly as the constant recirculation-based heat leak vanishes.

\subsubsection*{Thermal leakage}
To keep the system cold, a series of constant and flow-dependent heat leaks due to radiation, conduction and convection processes need to be compensated. The constant heat leak at zero flow is only due to losses at the cryostat and cold head and can be calculated from the heater output power if the cooling power of the PTR is known at the respective cold-head set-point. When recirculation is started, the umbilical and the heat exchangers are cooled down as well, even at very low flow. This represents another source of constant heat leak. It can be quantified from the offset of the power-axis intercept of the fit in figure~\ref{fig:gas_enthalpy_contributions_heater_power}, right, and the actual measurement of the required zero-flow heater power. We obtained the estimate $\SIrange{14}{34}{W}$ and conclude that the total constant heat leak is the sum of the constant heat leaks of the cryostat, the cold head, as well as the heat exchanger and the umbilical. The latter two only play a role when recirculating. In addition to the constant heat leaks, flow-dependent losses emerge when recirculating. One source of such a loss comes from the small inefficiency of the heat exchange. Moreover, flow-dependent temperature changes of the setup change the radiative heat input, the impact of heat conduction processes through holding structures and the convection by a change of the quality of the insulation vacuum. The sum of all these losses as a function of the flow is obtained from the heater output measurement. The temperature difference of the pre-cooler inlet and post-heater outlet allows us to calculate independently an efficiency from the enthalpy differences. We note that this efficiency is not equal to the efficiency of the heat exchange. This becomes obvious when considering the situation in which the cold outgoing gas from the return line would exchange its heat with a heat bath of the environment that provides the energy to heat up the gas, while the ingoing gas on the supply line enters the cryostat at room temperature. The temperature difference of gas inlet and outlet is zero, thus we could deduce a perfect heat exchange. For this reason, this method does not allow us to disentangle the heat exchange inefficiency and any other flow-dependent heat leaks from the heat exchanger and umbilical section. However, this measurement is not sensitive to flow-dependent heat leaks of the cryostat. Hence, the difference of the loss calculation of the heater power output and the gas inlet/outlet temperature measurement is equal to the flow-dependent heat leaks of the cryostat plus twice that of the heat exchanger, see figure~\ref{fig:he_efficiency_thermal_leakage}, right. We obtained a loss of $\SI{0.26}{W/slpm}$ from the sum of the inefficiency of the heat exchange and other flow-dependent losses on the heat exchangers and the umbilical, and can deduce that the latter are $< \SI{0.02}{W/slpm}$. Furthermore, we can conclude that the sum of the cryostat, heat exchanger and umbilical losses are $< \SI{0.04}{W/slpm}$.

\subsection{Slow control test}
The slow control system was subjected to a series of extensive tests to evaluate its reliability and performance.
The tests focused on benchmarking availability, effectiveness of the redundancy measures, the performance of the notification delivery system and system load under normal operations.

As described in section~\ref{sec:SC}, the SC system is designed to automatically react to single points of failure, this includes single services becoming unresponsive, network failures and failures that involve an entire server, for example, a hardware malfunction. 
In such cases Kubernetes ensures a redundancy switch, whereby all software services are migrated to a second stand-by server. 
We simulated severe server and network failures by turning off machines and disconnecting network cables; in all cases a full redundancy switch was automatically in effect after $\sim \SI{6}{min}$, as configured.
Thanks to this feature and to the micro-service architecture, over the past six months we have achieved an overall data collection availability of $\SI{99.6}{\%}$ (about $\SI{17}{hours}$ downtime in total), including scheduled and unscheduled maintenance. 
This demonstrates that the availability goal of $\SI{99.99}{\%}$ is realistic and within reach. We note that, for a large fraction of this time interval, the slow control system was still under design and testing phase.

The notification system proved to be extremely reliable: more than 1000 notifications were sent over a period of six months, of which $\SI{99.4}{\%}$ were successfully and timely delivered, with all undelivered messages being related to client-side issues. Furthermore, since all alarm messages are sent via two independent notification methods, email and SMS, it is unlikely that an alarm would remain undelivered even in the case of a client-side error.

We evaluated the system load under nominal operating conditions, which corresponds to two operators actively  monitoring and/or controlling the system.
When all services are running on a single server, a minimum of $\sim \SI{2}{GB}$ of memory are required for the system to perform well (without considering the operative system overhead). In nominal operation conditions an average network throughput of $\sim \SI{200}{kB/s}$ is observed, which is well within the capabilities of each server. 
Based on these findings, we estimate that the system can potentially serve up to about 200~users at the same time, where the limit is mainly due to the University's network average throughput capabilities.

\section{Summary and outlook}
\label{sec:outlook}

We designed and built a full-scale DARWIN demonstrator facility in the vertical dimension,  Xenoscope. The main objective of Xenoscope is to demonstrate for the first time electron drift in a LXe TPC over a distance of \SI{2.6}{m}. The facility will be a platform for testing several key technologies necessary to the realisation of the proposed DARWIN experiment.

We have shown the nominal operational reach of the Xenoscope facility by performing the first successful LXe filling and recuperation of \SI{80.66}{kg} of xenon. This operation allowed us to benchmark the components of the cryogenic system and to demonstrate the reliable and continuous operation of the purification loop. We performed a characterisation of the heat exchanger system and estimated the cooling power of the cryogenic system. Xenon extracted from the liquid phase was recirculated between 5 and 80~slpm in the purification loop, with a cryostat pressure range of \SIrange{1.77}{2.06}{bar}, at a cold head temperature of \SI{174.25}{K}. We demonstrated the cryopumping capability of the GXe storage array by recuperating the xenon over two days. We have also constructed a gravity-assisted LXe recuperation system, BoX, which will allow for faster xenon recuperation in liquid mode. This recuperation scheme will decrease the recuperation time for the full \SI{350}{kg} of xenon from the projected \SI{7.5}{days} in gas mode to less than one day. We have also demonstrated the reliability of our new slow control system. The monitoring and alarming system operated seamlessly throughout the commissioning phase, showing promising first steps in the use of open-source software, which may in the future be applicable to the DARWIN experiment.

Following the commissioning of the facility without an inner detector, the commissioning of the \SI{525}{mm} PM inside the cryostat is forthcoming, followed by the \SI{1}{m} TPC phase and finally the \SI{2.6}{m} tall TPC. The electron production from photoelectric effect on the surface of the photocathode has been successfully observed in a test setup. The charge collection readout was shown to have sufficient amplification and linearity to observe signals above noise level, in vacuum and LXe, and was reliably operated at temperatures around \SI{176}{K}. The PM and TPC field cage, the MPPC array and the liquid level control system are in production.

While the baseline HV distribution system of Xenoscope is an \mbox{off-the-shelf,} ceramic \sloppy{feedthrough} entering the LXe from the bottom, the facility was designed to test HV feedthrough prototypes entering through the top flange. These alternative options, currently under investigation, can be constructed with low-radioactivity materials and therefore be suitable for DARWIN.

In addition to demonstrating electron drift over \SI{2.6}{m}, our near-future plans with Xenoscope include the measurement of the longitudinal and transverse diffusion of  electrons emitted from the photocathode. The longitudinal diffusion affects the width of the signals, with wider signals for larger drift distances. This affects the capability of the TPC to resolve multiple scatters along the $z$-coordinate, which is imperative in rare event searches to distinguish background-like events, likely to produce multiple scatters, from signal-like events which are expected to interact only once. The transverse diffusion, on the other hand, affects the $x-y$ resolution of the detector. Previous measurements where performed over short drift lengths ($<\SI{20}{cm}$) and feature large systematic errors~\cite{Hogenbirk:2018knr,Albert:2016bhh}, revealing the need of new measurements for tonne-scale LXe TPCs, where charge diffusion is more relevant given the longer drift lengths. Xenoscope will provide the appropriate conditions to perform diffusion measurements at different drift fields for lengths up to \SI{2.6}{m}.

We also plan to carry out a campaign of measurements to determine optical properties of LXe. The working principle of a LXe TPC relies on the detection of  VUV photons, assuming that the xenon is transparent to its own scintillation light. However, light can be attenuated while traveling towards the photosensors, by absorption on impurities or due to Rayleigh scattering, affecting the reconstruction of the signals and the attainable energy threshold and resolution. Current LXe experiments assume a Rayleigh scattering length of \SI{30}{cm} and absorption length of \SI{50}{m}~\cite{Aprile:2015uzo}. These are both conservative values. The former is based on calculations~\cite{Seidel:2001vf}, and the latter on the assumption of a sub-ppb level concentration of water and a scaling of the results found in~\cite{Baldini:2004ph}, where an absorption length $>\SI{1}{m}$ was obtained for a \SI{100}{ppb} water concentration. Given the full length of Xenoscope, we will have the unique opportunity to improve the understanding of these two LXe properties, as well as the measurement of the refractive index and the group velocity for length scales that are relevant for the DARWIN experiment. In addition, we will be able to monitor these parameters as a function of temperature, pressure and xenon purity, including its water content.

In view of testing other technologies for DARWIN, we will evaluate the performance of new $2\times \SI{2}{in^2}$ square PMTs from Hamamatsu as well as low dark count VUV SiPMs in Xenoscope's TPC. The facility will also be  available to other members of the DARWIN collaboration to test relevant technologies, e.g.~bubble-assisted liquid-hole multipliers~\cite{Erdal:2019dkk}, hybrid photosensor solutions such as ABALONE~\cite{Ferenc:2017emv}, and Vacuum Silicon Photomultiplier Tubes~\cite{Barbarino:2014hka}, in conditions similar to those of the DARWIN observatory environment.

\acknowledgments
This work was supported by the European Research Council (ERC) under the European Union’s Horizon 2020 research and innovation programme, grant agreement No. 742789 (Xenoscope), by the SNF Grant 200020-188716 and by the University of Zurich. We thank Reto Maier and the mechanical workshop for their technical support and fabrication of the mechanical components. We also thank David Wolf and Achim Vollhardt of the electronics workshop for the design and realisation of the electrical systems. We thank Marta Gibert and Jonathan Spring for their help with the production of the photocathodes. We thank Guillaume Plante, Julien Masbou and Luca Scotto Lavina for the fruitful discussions regarding liquid xenon recovery.
\newpage
\bibliographystyle{jhep}
\bibliography{main-file}

\end{document}